\documentclass[ALICE,manyauthors]{cernphprep}

\usepackage[comma,square,numbers,sort&compress]{natbib}
\usepackage{hyperref}
\usepackage{lineno}
\usepackage{upgreek}
\usepackage[graphicx]{realboxes}
\usepackage[T1]{fontenc}
\newcommand{\krl}{\ensuremath{\kern-0.18em}}
\newcommand{\krr}{\ensuremath{\kern-0.09em}}
\newcommand{\tms}{\ensuremath{\kern-0.1em\times\kern-0.2em}}
\newcommand{\ptt}{\ensuremath{p_{\mathrm{T}}}\xspace}
\newcommand{\mT}{\ensuremath{m_{\mathrm{T}}}\xspace}

\newcommand{\pb}{Pb--Pb\xspace}

\newcommand{\ppb}{p--Pb\xspace}

\newcommand{\ada}{A--A\xspace}

\newcommand{\rs}[1][13~TeV]{\ensuremath{\sqrt{s}=}~#1\xspace}

\newcommand{\rsnn}[1][2.76~TeV]{\ensuremath{\sqrt{s_{\mathrm{NN}}}=}~#1\xspace}

\newcommand{\igz}{\ensuremath{\mathrm{INEL}\krl>\krr0}\xspace}
\newcommand{\gvc}{\ensuremath{\mathrm{GeV}\krl/\krr c}\xspace}
\newcommand{\gvcc}{\ensuremath{\mathrm{GeV}\krl/\krr c^{2}}\xspace}

\newcommand{\mvcc}{\ensuremath{\mathrm{MeV}\krl/\krr c^{2}}\xspace}

\newcommand{\pion}{\ensuremath{\uppi}\xspace}
\newcommand{\pix}{\ensuremath{\pion^{\pm}}\xspace}

\newcommand{\kk}{\ensuremath{\mathrm{K}}\xspace}
\newcommand{\kx}{\ensuremath{\mathrm{K}^{\pm}}\xspace}
\newcommand{\km}{\ensuremath{\mathrm{K}^{-}}\xspace}
\newcommand{\kp}{\ensuremath{\mathrm{K}^{+}}\xspace}
\newcommand{\kz}{\ensuremath{\mathrm{K}^{0}_{\mathrm{S}}}\xspace}

\newcommand{\ks}{\ensuremath{\mathrm{K^{*0}}}\xspace}

\newcommand{\ksm}{\ensuremath{\mathrm{K^{*}\krr(892)^{0}}}\xspace}
\newcommand{\ksbm}{\ensuremath{\mathrm{\overline{K}^{*}\!(892)^{0}}}\xspace}

\newcommand{\ph}{\ensuremath{\upphi}\xspace}
\newcommand{\phm}{\ensuremath{\ph(1020)}\xspace}

\newcommand{\ksk}{\ensuremath{\ks\krl/\mathrm{K}}\xspace}

\newcommand{\kskz}{\ensuremath{\ks\krl/\kz}\xspace}

\newcommand{\pks}{\ensuremath{\mathrm{p}\kern-0.1em/\ks}\xspace}
\newcommand{\pksm}{\ensuremath{\mathrm{p}\kern-0.1em/\ksm}\xspace}

\newcommand{\phipi}{\ensuremath{\ph\krl/\krr\pion}\xspace}

\newcommand{\phik}{\ensuremath{\ph\krl/\mathrm{K}}\xspace}

\newcommand{\phikz}{\ensuremath{\ph\krl/\kz}\xspace}

\newcommand{\pphi}{\ensuremath{\mathrm{p}\kern-0.1em/\krl\ph}\xspace}
\newcommand{\pphim}{\ensuremath{\mathrm{p}\kern-0.1em/\krl\phm}\xspace}

\newcommand{\xiphi}{\ensuremath{\Xi\krl/\krr\ph}\xspace}

\newcommand{\omphi}{\ensuremath{\Omega\kern-0.05em/\krr\ph}\xspace}
\newcommand{\omxphi}{\ensuremath{(\Omega^{-}+\overline{\Omega}^{+})\kern-0.05em/\krr\ph}\xspace}
\newcommand{\ommphi}{\ensuremath{\Omega^{-}\kern-0.05em/\krr \ph}\xspace}
\newcommand{\ompphi}{\ensuremath{\overline{\Omega}^{+}\kern-0.05em/\krr \ph}\xspace}
\newcommand{\omphim}{\ensuremath{\Omega\kern-0.05em/\krr\phm}\xspace}

\newcommand{\xipi}{\ensuremath{\Xi\kern-0.1em/\krr\pion}\xspace}
\newcommand{\ompi}{\ensuremath{\Omega\kern-0.05em/\krr\pion}\xspace}

\newcommand{\dd}{\ensuremath{\mathrm{d}}}
\newcommand{\mpt}{\ensuremath{\langle\ptt\rangle}\xspace}

\newcommand{\dndy}{\ensuremath{\dd N\krl/\krr\dd y}\xspace}

\newcommand{\dnc}{\ensuremath{\langle\dd N_{\mathrm{ch}}\kern-0.06em /\kern-0.13em\dd\eta\rangle_{|\eta|<0.5}}\xspace}

\newcommand{\dedx}{\ensuremath{\dd E\krl/\krr\dd x}\xspace}
\newcommand{\stpc}{\ensuremath{\sigma_{\mathrm{TPC}}}\xspace}
\newcommand{\stof}{\ensuremath{\sigma_{\mathrm{TOF}}}\xspace}

\newcommand{\mpik}{\ensuremath{m_{\pion\mathrm{K}}}\xspace}
\newcommand{\mkk}{\ensuremath{m_{\mathrm{KK}}}\xspace}

\begin{document}%

%%%%%%%%%%%%%%%  Title page %%%%%%%%%%%%%%%%%%%%%%%%
\begin{titlepage}
% CERN-EP-2019-245
\PHyear{2019}
\PHnumber{245}      % required, will be obtained from PH
\PHdate{28 October}
%

%%% Put your own title + short title here:
\title{Multiplicity dependence of K*(892)$^{\mathbf{0}}$ and $\boldsymbol{\upphi}$(1020)\\ production in pp collisions at $\boldsymbol{\sqrt{s}}$~=~13~TeV}
\ShortTitle{\ks and \ph in pp collisions vs. multiplicity}   % appears on right page headers

%%% Do not change the next lines
\Collaboration{ALICE Collaboration\thanks{See Appendix~\ref{app:collab} for the list of collaboration members}}
\ShortAuthor{ALICE Collaboration} % appears on left page headers, do not change

\begin{abstract}
Measurements of identified hadrons as a function of the charged-particle multiplicity in pp collisions enable a search for the onset of collective effects in small collision systems. With such measurements, it is possible to study the mechanisms that determine the shapes of hadron transverse momentum (\ptt) spectra, to search for possible modifications of the yields of short-lived hadronic resonances due to scattering effects in the hadron-gas phase, and to investigate different explanations for the multiplicity evolution of strangeness production provided by phenomenological models. In this paper, these topics are addressed through measurements of the \ksm and \phm mesons at midrapidity in pp collisions at \rs as a function of the charged-particle multiplicity. The results include the \ptt spectra, \ptt-integrated yields, mean transverse momenta, and the ratios of the yields of these resonances to those of longer-lived hadrons. Comparisons with results from other collision systems and energies, as well as predictions from phenomenological models, are also discussed.
\end{abstract}
\end{titlepage}
\setcounter{page}{2}

\section{Introduction\label{sec:intro}}

At the LHC, recent studies of \ppb and pp collisions with high charged-particle multiplicities have shown striking similarities to \pb collisions. Measurements of azimuthal correlations of particles and anisotropic flow ($v_{2}$)~\cite{ALICE_flow_2014,ALICE_flow_2019,CMS_v2_pp,ALICE_ridge_pPb5,CMS_v2_pPb5,ATLAS_v2_pp13,ATLAS_flow_2017} suggest the possibility of collective effects in small collision systems. However, the origins of these effects is not yet fully understood and it remains an open question whether the underlying causes are the same as in large collision systems such as \pb and Xe--Xe. In order to investigate the origin of these effects, the ALICE Collaboration has measured the \ptt spectra and total yields of identified hadrons in \ppb collisions as a function of the charged-particle multiplicity~\cite{ALICE_piKpLambda_pPb5,ALICE_Kstar_phi_pPb5,ALICE_Xi_Omega_pPb5,ALICE_Sigmastar_Xistar_pPb5}, which is used as a measure of the ``activity" of the event. The ALICE Collaboration has also studied the multiplicity dependence of light-flavor hadron production in pp collisions for many species: for \pix, \kx, \kz, \ksm, p, \phm, $\Lambda$, $\Xi^{-}$, $\Omega^{-}$, and their antiparticles at \rs[7~TeV]~\cite{ALICE_strangeness_pp7,ALICE_LF_pp7} and for \kz, $\Lambda$, $\Xi^{-}$, $\Omega^{-}$, and their antiparticles at \rs~\cite{ALICE_strange_pp13}.
%For pp collisions, the ALICE Collaboration has conducted studies of the production of many light-flavor hadrons (\pix, \kx, \kz, \ksm, p, \phm, $\Lambda$, $\Xi^{-}$, $\Omega^{-}$, and their antiparticles) at \rs[7~TeV]~\cite{ALICE_strangeness_pp7,ALICE_LF_pp7} and a study of strange-hadron production (\kz, $\Lambda$, $\Xi^{-}$, $\Omega^{-}$, and their antiparticles) in pp collisions at \rs~\cite{ALICE_strange_pp13}, both as a function of the charged-particle multiplicity.
This paper reports on an extension of these studies: a measurement of the multiplicity evolution of the production of \ksm, \ksbm, and \phm mesons in pp collisions at \rs, the highest energy reached by the LHC in runs 1 and 2. The present study takes advantage of a pp data set recorded during Run 2 of the LHC in 2015 with an integrated luminosity of 0.88~nb$^{-1}$.
%, from which a measurement of \ks and \ph mesons in inelastic pp collisions (not divided into multiplicity classes) at \rs has already been reported~\cite{ALICE_LF_pp13}.
For the remainder of this paper, the average of \ksm and \ksbm will be denoted as \ks, while the \phm will be denoted as \ph.

The ratios of the yields of strange hadrons to pion yields are observed to be enhanced in nucleus--nucleus (\ada) collisions relative to minimum bias pp collisions~\cite{NA57_strange,STAR_strange_AuAu200,ALICE_multistrange_PbPb276}, with the yields in central \ada collisions being well described by statistical thermal models~\cite{Cleymans_thermal_2006,Andronic_thermal_2006,Andronic_thermal_2011,Andronic_thermal_2018}. At the LHC, these ratios are observed to increase with the charged-particle multiplicity in pp and \ppb collisions~\cite{ALICE_strangeness_pp7,ALICE_LF_pp7,ALICE_strange_pp13,ALICE_piKpLambda_pPb5,ALICE_Kstar_phi_pPb5,ALICE_Xi_Omega_pPb5};  the magnitude of the change from low to high multiplicity increases with the strangeness content of the hadron. The ratios in high-multiplicity pp and \ppb collisions reach the values observed in peripheral \pb collisions and generally follow similar trends as the multiplicity increases from pp to p--A to \ada collisions. Furthermore, the yields of strange particles are consistent between \rs[7] and 13~TeV for similar charged-particle multiplicities. These results suggest that the yields of these hadrons depend primarily on the charged-particle multiplicity and are independent of the collision system and energy.

Several theoretical explanations of the multiplicity evolution of strange-hadron production have been put forward, including canonical suppression, rope hadronization, and core-corona effects. In statistical thermal models of large collision systems, strangeness production is described through the use of a grand canonical ensemble, where strangeness conservation is realized on average across the volume of the system. In the canonical suppression picture, strangeness production in small systems is instead described using a canonical ensemble, requiring the exact local conservation of strangeness within the small volume~\cite{Hamieh_canonical,Cleymans_canonical_2004,ALICE_LF_pp7}. As the size of the system decreases, it makes a transition from the grand-canonical to the canonical description, leading to a decrease in strange-hadron yields with decreasing multiplicity. In the rope-hadronization picture, the larger and denser collision systems form color ropes~\cite{Biro_ColorRopes,Bialas_1985,Armesto_1995}, groups of overlapping strings that hadronize with a larger effective string tension. This effect, implemented in models such as DIPSY~\cite{preDIPSY,DIPSY,DIPSY_ColorRopes}, also leads to an increase in the production of strange hadrons with increasing charged-particle multiplicity. Core-corona separation is implemented in a variety of models, including EPOS~\cite{Drescher_GribovRegge,EPOS_2010,EPOS3,EPOS_LHC} and those described in~\cite{Kanakubo_2018,Akamatsu_2018}. In these models, the part of the collision system that has high string or parton densities becomes a ``core" region that may evolve as a quark--gluon plasma; this is surrounded by a more dilute ``corona" for which fragmentation occurs as in the vacuum. Strangeness production is higher in the core region, which makes up a greater fraction of the volume of the larger collision systems. This also results in strangeness enhancement with increasing multiplicity.

The \ph meson is a useful probe for the study of strangeness enhancement. The \ph contains two strange valence (anti)quarks, but has no net strangeness. Its production should therefore not be canonically suppressed, while the production of hadrons with open strangeness (\textit{e.g.} kaons or $\Xi$) may be canonically suppressed~\cite{ALICE_LF_pp7}. It has, in fact, been rather difficult to describe enhancement of \ph-meson production in a framework that involves canonical suppression~\cite{ALICE_LF_pp7}. In contrast, in the rope-hadronization or core-corona interpretations, the yields of \ph mesons evolve with multiplicity similarly to particles with open strangeness, leading to an expected increase in the \ptt-integrated \phipi ratio with increasing charged-particle multiplicity. Measurements of \ph-meson production as a function of the multiplicity may help to distinguish between the various explanations of strangeness enhancement in small systems.

One of the main motivations for studying resonances like \ks and \ph in heavy-ion collisions is to learn more about the properties (temperature and lifetime) of the hadronic phase of the collision. When short-lived resonances (such as $\uprho(770)^{0}$, \ks, and $\Lambda(1520)$) decay, their daughters may re-scatter in the hadronic phase, leading to a reduction in the measurable resonance yields; conversely, resonances may also be regenerated due to quasi-elastic scattering of hadrons through a resonance state~\cite{Bleicher_Stoecker,Torrieri_2001,Markert_2008,Vogel_2010,Bliecher_Aichelin,EPOS_resonances_PbPb}. Centrality-dependent suppression of $\uprho(770)^{0}$, \ks, and $\Lambda(1520)$ production was observed in \pb collisions~\cite{ALICE_rho_PbPb276,ALICE_Kstar_phi_PbPb276,ALICE_Kstar_phi_PbPb276_highpt,ALICE_Lambda1520_PbPb276}, and a hint of suppression of \ks suppression was reported for \ppb collisions~\cite{ALICE_Kstar_phi_pPb5}. Observations of a similar suppression in high-multiplicity pp collisions (e.g., the \ksk ratio in pp collisions at \rs[7~TeV]~\cite{ALICE_LF_pp7}) might be an indication for a hadronic phase with non-zero lifetime in high-multiplicity pp collisions.

It was observed that the slopes of hadron \ptt spectra increase with increasing multiplicity in pp and \ppb collisions~\cite{ALICE_LF_pp7,ALICE_strange_pp13,ALICE_piKpLambda_pPb5,ALICE_Kstar_phi_pPb5}. This is at least qualitatively similar to the behavior observed in \pb collisions, where the observed increase in the slopes can be attributed to a collective expansion of the system; low-\ptt particles receive a radial momentum boost, which is greater in higher multiplicity collisions~\cite{ALICE_piKp_PbPb,ALICE_Kstar_phi_pPb5}. The color reconnection (CR) mechanism~\cite{Gustafson_1988,Gustafson_1994,PYTHIA6o4,Ortiz_2013,Bierlich_2015} can lead to collective flow-like effects, even in small collision systems and in event generators like PYTHIA that do not include QGP formation. The increase in the slopes of the \ptt spectra is also mirrored in the trend of the measured mean transverse momenta \mpt. In contrast to the yields, which evolve along a continuous trend with multiplicity across different collision systems, the \mpt values of light-flavor hadrons follow different trends in pp, \ppb, and \pb collisions~\cite{ALICE_piKpLambda_pPb5,ALICE_Kstar_phi_pPb5,ALICE_strangeness_pp7,ALICE_piKp_PbPb}, with a faster increase for the smaller systems. The \mpt values in the highest multiplicity pp collisions reach, or in some cases exceed, the \mpt values observed in central \pb collisions. The increase in \mpt in pp collisions is due to changes in the shapes of the \ptt spectra at low \ptt; for $\ptt\gtrsim4$~\gvc, the shapes of hadron \ptt spectra are essentially independent of multiplicity~\cite{ALICE_charged_pp5_13,ALICE_strange_pp13}.

The results reported here will allow the study of \ks and \ph production as functions of both energy and multiplicity in pp collisions. The presented results reach higher values of multiplicity than previously measured in pp collisions and therefore provide important additional information on the production of light-flavor hadrons at LHC energies. This paper is organized as follows. The ALICE detector and the criteria adopted for data selection are described in Section~\ref{sec:selection}. A summary of the data analysis procedure is given in Section~\ref{sec:analysis}. The results are presented and discussed in Section~\ref{sec:results}, followed by a summary and conclusions in Section~\ref{sec:conclusions}.

\section{Event and Track Selection\label{sec:selection}}

The ALICE detector is described in detail in~\cite{ALICE_detector,ALICE_performance_2014}. The sub-detectors that are relevant to the analysis described in this paper are the Time Projection Chamber (TPC), the Time-of-Flight detector (TOF), the Inner Tracking System (ITS), the V0 detectors, and the T0 detector. The TPC and ITS are used for tracking and finding the primary vertex, while the TPC and TOF are used for particle identification. The V0 detectors are used for triggering and to define the multiplicity estimator at forward rapidities (pseudorapidity ranges $-3.7<\eta<-1.7$ and $2.8<\eta<5.1$). The T0 detector is used for triggering and to provide timing information (including a start signal for the TOF).

The \ks and \ph mesons are reconstructed from a sample of $5\tms10^{7}$ pp collisions at \rs recorded in 2015. The minimum bias trigger required hits in both V0 detectors in coincidence with proton bunches arriving from both directions. Beam-induced background and pile-up events are removed \mbox{offline;} see~\cite{ALICE_strange_pp13,ALICE_performance_2014} for details. Selected events must also have a primary collision vertex reconstructed with the two innermost layers of the ITS and located within $\pm10$~cm along the beam axis of the nominal center of the ALICE detector. Results in this paper are presented for different event classes corresponding to subdivisions of the ``\igz" event class, which is defined as the set of inelastic collisions with at least one charged particle in the range $|\eta|<1$~\cite{ALICE_multiplicity_13}. The \igz sample is divided into multiplicity classes based on the total charge deposited in both V0 detectors (called the ``V0M amplitude"). Thus, the event classes are determined by the number of charged particles at forward rapidities, while the \ks and \ph yields are measured at midrapidity $(|y|<0.5)$; this is to avoid correlations between the \ks and \ph yields and the multiplicity estimator. Particle yields, yield ratios, and mean transverse momenta are plotted for different multiplicity classes (which correspond to different centralities for \ada collisions) as functions of the mean charged-particle multiplicity density at midrapidity $\dnc$, where $\eta$ is the pseudorapidity in the lab frame. As in~\cite{ALICE_strange_pp13}, the various multiplicity classes are denoted using Roman numerals, with class I (X) having the highest (lowest) multiplicity. See Table~\ref{table:mult} for the values of \dnc measured for each V0M multiplicity class.

Since the \ks and \ph mesons are short-lived (i.e., their lifetimes are of the order of $\sim10^{-23}$~s and their decay vertices cannot be distinguished from the primary collision vertex), they cannot be measured directly by the detector. Instead, they are reconstructed via their hadronic decays to charged pions and kaons: $\ks\rightarrow\pix\kk^{\mp}$ (branching ratio $66.503\pm0.014$\%) and $\ph\rightarrow\kp\km$ (branching ratio $49.2\pm0.5$\%)~\cite{PDG}. Charged tracks are selected using a set of standard track-quality criteria, described in detail in~\cite{ALICE_Kstar_phi_pPb5}. Pions and kaons are identified using the specific ionization energy loss \dedx measured in the TPC and the flight time measured in the TOF. Where the \dedx resolution of the TPC is denoted as \stpc, pions and kaons are required to have \dedx values within 2\stpc of the expected value for $p>0.4$~\gvc, within 4\stpc for $0.3<p<0.4$~\gvc, and within 6\stpc for $p<0.3$~\gvc (typically, $\stpc\sim5\%$ of the measured \dedx value). When a pion or kaon track is matched to a hit in the TOF, the time-of-flight value is required to be within 3\stof of the expected value ($\stof\sim80$~ps)~\cite{ALICE_timing}. These event- and track-selection criteria are varied from their default values and the resulting changes in the yields are incorporated into the systematic uncertainties, which are summarized in Table~\ref{table:sys}.

\begin{table}
\caption{Charged-particle multiplicity densities at midrapidity \dnc for the \igz class and the various V0M multiplicity classes~\cite{ALICE_strange_pp13}.}
\begin{center}
\begin{tabular}{ c c }
\hline
Class & \dnc\\\hline
\igz & 6.89$\pm$0.11 \\
I & 25.75$\pm$0.40 \\
II & 19.83$\pm$0.30 \\
III & 16.12$\pm$0.24 \\
IV & 13.76$\pm$0.21 \\
V & 12.06$\pm$0.18 \\
VI & 10.11$\pm$0.15 \\
VII & 8.07$\pm$0.12 \\
VIII & 6.48$\pm$0.10 \\
IX & 4.64$\pm$0.07 \\
X & 2.52$\pm$0.04 \\\hline
\end{tabular}
\end{center}
\label{table:mult}
\end{table}

\section{Data Analysis\label{sec:analysis}}

The \ks and \ph signals are extracted using the same invariant mass reconstruction method described in~\cite{ALICE_Kstar_phi_PbPb276,ALICE_Kstar_phi_pPb5}. Invariant mass distributions of unlike-charge $\pion$K or KK pairs in the same event are reconstructed after particle identification. The combinatorial background is estimated using multiple methods. In the ``like-charge" method, tracks of identical charge from the same event are combined to form pairs. This background is $2\sqrt{n_{- -}n_{++}}$, where $n_{- -}$ and $n_{++}$ are the number of negative-negative and positive-positive pairs in each invariant mass bin, respectively. In the ``mixed-event" method, tracks from one event are combined with oppositely charged tracks from up to 5 other events with similar primary vertex positions and multiplicity percentiles. Specifically, it is required that the longitudinal positions of the primary vertices differ by less than 1~cm and the multiplicity percentiles computed using the V0M amplitude differ by less than 5\%. The mixed-event $\pion$K (KK) background is normalized so that it has the same integral as the unlike-charge same-event distribution in the invariant mass range $1.1<\mpik<1.15$~\gvcc ($1.05<\mkk<1.08$~\gvcc). In evaluating the systematic uncertainties, the boundaries of the normalization region for the mixed-event background are varied by $\sim100$~\mvcc for the \ks analysis and $\sim10$~\mvcc for \ph.

After subtraction of the combinatorial background, the invariant mass distribution consists of a resonance peak sitting on top of a residual background of correlated pairs. This correlated background contains contributions from jets, resonance decays in which a daughter is misidentified, and decays with more than two daughters. In the analysis of the \ph meson in pp collisions, the signal-to-background ratio is large and the background is observed to vary slowly in the region of the peak. For these reasons, a third approach is also used to describe the background in the \ph analysis; the combinatorial background is not subtracted, but is instead parameterized together with the residual background using a function as described below. This has the advantage of providing smaller statistical uncertainties than the other methods.

For $\ptt<4$~\gvc, all three methods provide good descriptions of the KK background and give \ph yields within a few percent of each other. The final \ph yields for $\ptt<4$~\gvc are the averages of those extracted using the three methods of describing the combinatorial background, while the spread among the results for the different methods is incorporated into the systematic uncertainties. As \ptt increases, the yields of hadrons decrease, along with the magnitudes of all of the combinatorial backgrounds studied. The mixed-event background, which lacks any contribution from correlated pairs, is observed to become smaller than the same-event (like- or unlike-charge) combinatorial backgrounds as \ptt increases, eventually tending to 0 for \ptt values higher than the ranges considered here. While the mixed-event background could still be used for the \ph analysis for $4\leq\ptt\leq8$~\gvc, the two other techniques have smaller statistical fluctuations in this \ptt range. Consequently, the mixed-event technique is not used for the analysis of \ph for $\ptt>4$~\gvc. The mixed-event technique is the primary method used for the extraction of the \ks yields; variations of the yield due to the use of a like-charge background are covered by the systematic uncertainties. However, for $\ptt<0.8$~\gvc in multiplicity class I, the like-charge method is preferred, since it provides a better description of the background. At high \ptt, the mixed-event  background for the \ks analysis exhibits the same behavior as for the \ph, but the problems appear at higher \ptt values than for \ph. The mixed-event technique therefore remains the best available option for this \ks analysis, even at the high end of the \ptt range that was studied.
%Since the mixed-event background is of decreased utility at high \ptt and since two other techniques describe the combinatorial background for the \ph analysis well, the mixed-event technique is not used for the analysis of \ph for $\ptt>4$~\gvc.
%The mixed-event technique introduces large statistical uncertainties at high \ptt and is therefore not used for $\ptt>4$~\gvc. (except event mixing at high \ptt)
%However, the mixed-event background for \ks remains useable out to higher \ptt than for \ph, covering the whole \ptt range studied here.

The invariant mass distributions are fitted with a peak function added to a smooth residual background function. For \ks, the peak is described using a Breit-Wigner function. The mass resolution of the detector for the $\ph\rightarrow\km\kp$ channel is of the same order of magnitude as the \ph width. Therefore, the \ph peak is described using a Voigt function: a convolution of a Breit-Wigner function and a Gaussian which accounts for the mass resolution of the detector. The \ks and \ph width parameters are by default fixed to their vacuum values; to calculate the systematic uncertainties, these parameters are allowed to vary freely and the \ph resolution parameter is fixed to the values (approximately 1--2~\mvcc) extracted from the Monte Carlo simulations described below. The residual background is parameterized using a second-order polynomial. To evaluate the systematic uncertainties in the \ks yields, a third-order polynomial is used instead. For the \ph systematic uncertainties, a first-order polynomial and a function of the form \mbox{$A+B\mkk+C\sqrt{\mkk-2M(\mathrm{K}^{\pm})}$} are used. Here, $A$, $B$, and $C$ are free parameters, \mkk is the kaon-kaon pair invariant mass, and $M(\mathrm{K}^{\pm})$ is the mass of the \kx. The fits are performed in the invariant mass intervals $0.75<\mpik<1.07$~\gvcc for the \ks analysis and $0.995<\mkk<1.09$~\gvcc for the \ph. The ranges of the fits are varied by $\sim20$~\mvcc for \ks and $\sim10$~\mvcc for \ph; the resulting changes in the yields are included in the systematic uncertainties. Finally, particle yields are extracted by integrating the invariant mass distribution in the peak region ($0.798\leq\mpik\leq0.994$~\gvcc for \ks and $1.01\leq\mkk\leq1.03$~\gvcc for \ph), subtracting the integral of the residual background function under the peak, and adding the yields in the tails of the peak fit function outside the integration region. The systematic uncertainty arising from ``signal-extraction", as quoted in Table~\ref{table:sys}, covers the aforementioned variations in the combinatorial background, mixed-event normalization region, residual background function, peak function, and fit range. An additional uncertainty originates from the procedure used to match track segments in the ITS with tracks in the TPC. The branching ratio correction for the \ph yield introduces a 1\% uncertainty, while the corresponding uncertainty for \ks is negligible. Uncertainties in the yields due to uncertainties in the material budget of the detector and the cross sections for hadronic interactions in that material are taken from a previous study~\cite{ALICE_Kstar_phi_pPb5}.

The raw particle yields are corrected for the branching ratios, as well as the acceptance and efficiency of the reconstruction procedure. This correction is calculated using several different event generators (PYTHIA6 Perugia 2011 tune~\cite{PYTHIA6_Perugia}, PYTHIA8 Monash 2013 tune~\cite{PYTHIA8_Monash}, and EPOS-LHC~\cite{EPOS_LHC}), with particles propagated through a simulation of the detector using GEANT3~\cite{GEANT3}. No dependence on the generator is observed and the average correction for the three generators is used in order to reduce statistical fluctuations. This correction is of the same order as reported in~\cite{ALICE_Kstar_phi_pPb5}. A dependence on multiplicity is observed; for $\ptt<3$~\gvc, the correction increases by $\sim10$\% from multiplicity class I to class X.
%, with a small dependence on multiplicity (a few percent, decreasing with increasing multiplicity).
A ``signal-loss" correction is also applied, which accounts for \ks and \ph mesons in non-triggered events. This is evaluated using the same simulations as the acceptance and efficiency. To calculate this correction factor, the simulated resonance \ptt spectrum before triggering and event selection is divided by the corresponding \ptt spectrum after those selections for each multiplicity class. The signal-loss correction typically deviates from unity by $<1\%$, but can deviate by $\sim10\%$ at low \ptt for the lowest multiplicity class. The PYTHIA6 simulation is used to obtain the central values for this correction, while an uncertainty is evaluated by comparing the central values to those given by PYTHIA8 and EPOS-LHC. Finally, the \ptt spectra are normalized by the number of accepted events and corrected as in~\cite{ALICE_strange_pp13} to account for \igz events that do not pass the event-selection criteria. This correction, which is calculated using the PYTHIA6 simulation, is most important (24\%) for the lowest multiplicity class and is $<1$\% for high-multiplicity collisions (classes I-VIII).
%The default value of this correction is calculated using the PYTHIA6 simulation, while the systematic uncertainties are evaluated by changing the event generator.

\begin{table}
\caption{Sources of systematic uncertainties for the \ptt spectra of \ks and \ph reported for low, intermediate, and high \ptt. When only one value is given for one particle, the uncertainty does not depend on \ptt. ``Signal extraction" includes variations of the combinatorial background, mixed-event normalization region, fitting region, peak shape, and residual background function. The ``signal-loss" uncertainty is multiplicity-dependent, hence values are quoted for the highest and lowest multiplicity classes (I and X, respectively). The text ``negl." indicates a negligible uncertainty and ``had. int. cross sec." is short for ``hadronic interaction cross section."}
\begin{center}
\begin{tabular}{| r | c c c | c c c |}
\hline
Particle & & \ks & & &\ph & \\
\ptt (\gvc) & 0.2 & 2.2 & 6.5 & 0.7 & 2 & 6 \\\hline
event/track selection & 4.3\% & 1.6\% & 2.9\% & 2.7\% & 2.9\% & 3.2\%\\
signal extraction & 10.3\% & 6.7\% & 7.7\% & 2.7\% & 3.1\% & 3.1\%\\
ITS-TPC matching & & 2.0\% & & & 2.0\% & \\
branching ratio & & negl. & & & 1.0\% & \\
material budget & 2.0\% & 0.5\% & negl. & 5.3\% & 1.0\% & negl.\\
had. int. cross sec. & 2.6\% & 1.2\% & negl. & 2.1\% & 2.6\% & negl.\\\hline
signal loss, class I & & negl. & & & negl. & \\
signal loss, class X & 3.9\% & 2.4\% & 0.9\% & 2.3\% & 4.8\% & 2.2\%\\\hline
%normalization, class I & & 1.7\% & & & 1.7\% & \\
%normalization, class X & & 2.3\% & & & 2.3\% & \\\hline
\end{tabular}
\end{center}
\label{table:sys}
\end{table}

\begin{figure*}
\includegraphics[width=19pc]{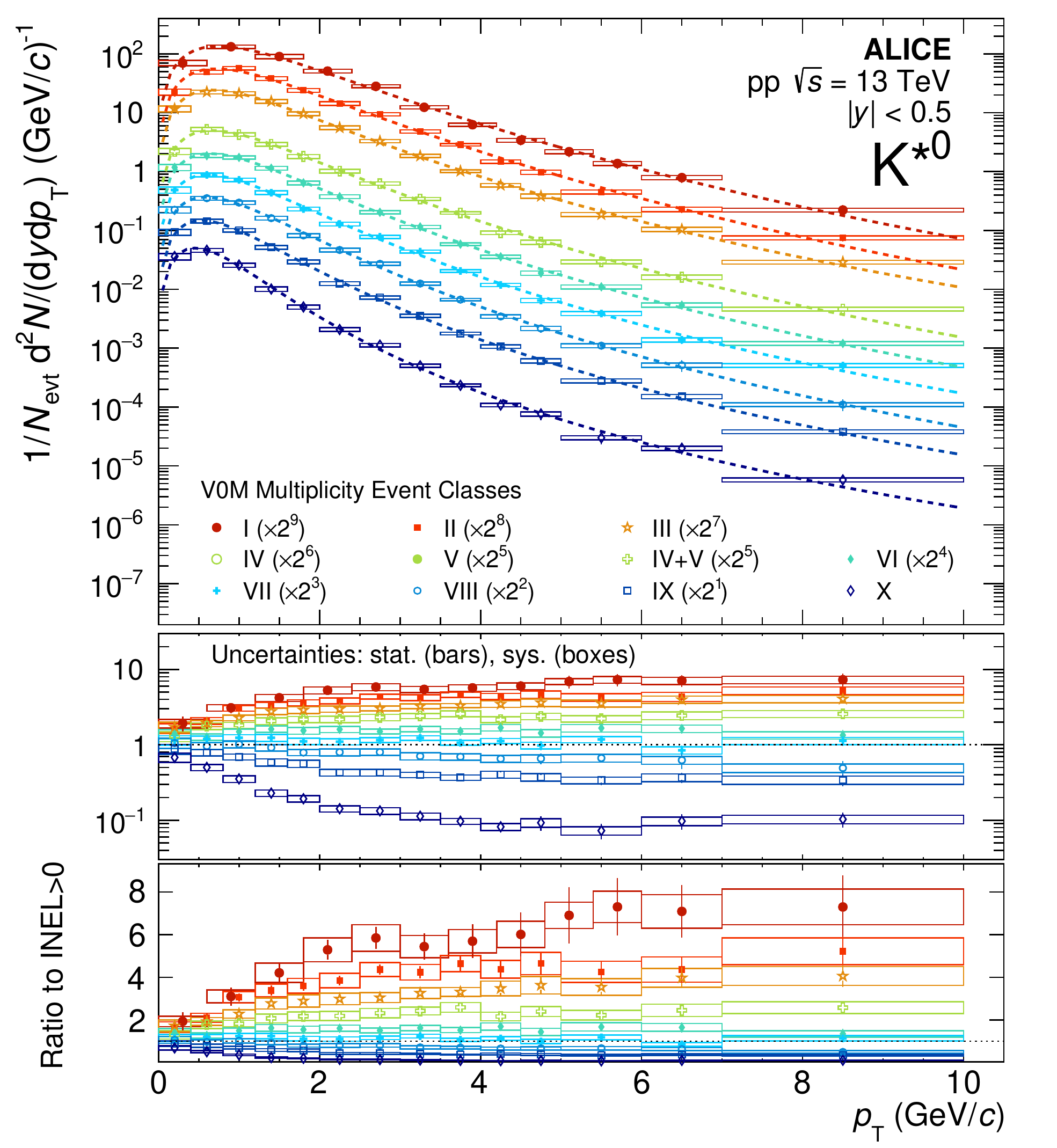}
\includegraphics[width=19pc]{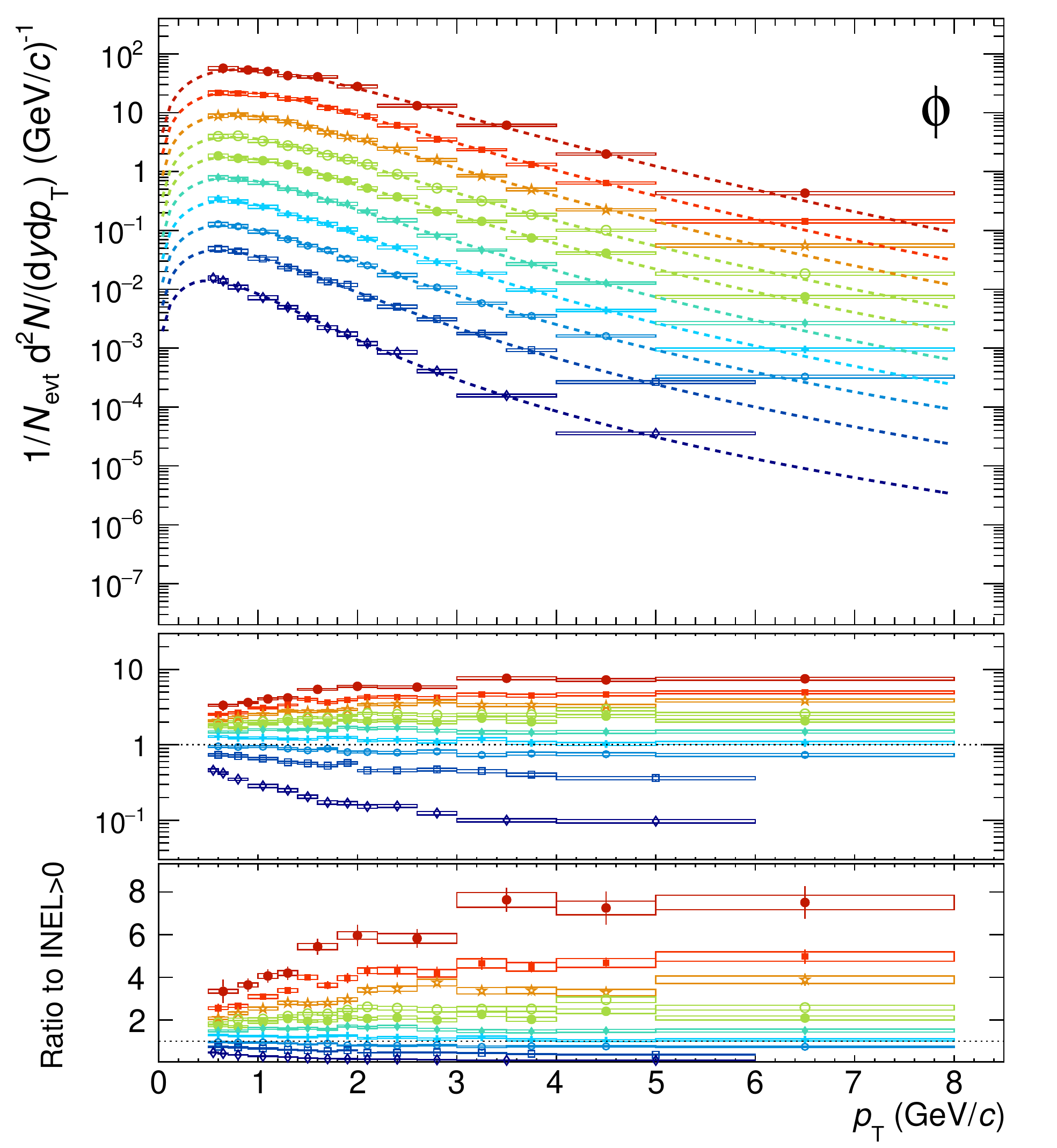}
\caption{\label{fig:spectra}(Color online)  \ptt spectra of \ks and \ph in pp collisions at \rs for different multiplicity classes, scaled by factors as indicated. The lower panels show the ratios of the multiplicity-dependent \ptt spectra to the multiplicity-integrated \igz spectra (with both linear and logarithmic vertical scales).}
\end{figure*}

\section{Results\label{sec:results}}

The \ptt spectra for \ks and \ph in the various multiplicity classes, as well as the ratios of these spectra to the inclusive \igz spectrum, are shown in Fig.~\ref{fig:spectra}. For $\ptt\lesssim4$~\gvc the increase in the slopes of the \ptt spectra from low to high multiplicity is clearly visible. For higher \ptt, the spectra in different multiplicity classes all have the same shape, indicating that the processes that change the shape of the \ptt spectra in different multiplicity classes are dominant primarily at low \ptt. A similar behavior was reported for unidentified charged hadrons, \kz, $\Lambda$, $\Xi$, and $\Omega$ for the same collision system~\cite{ALICE_charged_pp5_13,ALICE_strange_pp13}.

\begin{figure*}
\includegraphics[width=38pc]{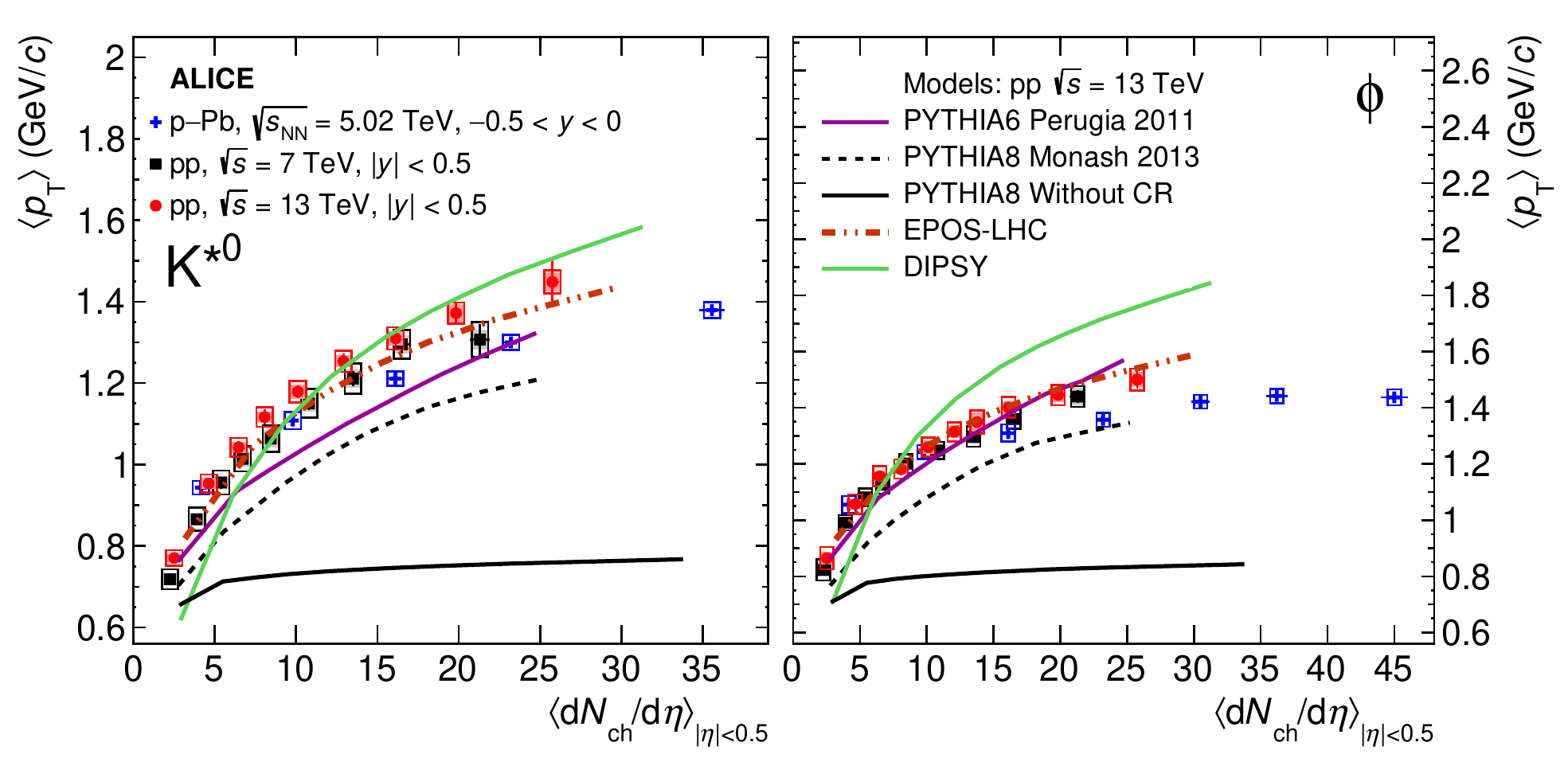}
\caption{\label{fig:mpt}(Color online) Mean transverse momenta \mpt of \ks and \ph as functions of \dnc. Results are shown for pp collisions at \rs[13] and 7~TeV~\cite{ALICE_LF_pp7}, as well as for \ppb collisions at \rsnn[5.02~TeV]~\cite{ALICE_Kstar_phi_pPb5}. The measurements in pp collisions at \rs are also compared to values from common event generators~\cite{PYTHIA6_Perugia,PYTHIA8_Monash,EPOS_LHC,DIPSY}. Bars represent statistical uncertainties, open boxes represent total systematic uncertainties, and shaded boxes show the systematic uncertainties that are uncorrelated between multiplicity classes (negligible for \ppb).}
\end{figure*}

\begin{figure*}
\includegraphics[width=19pc]{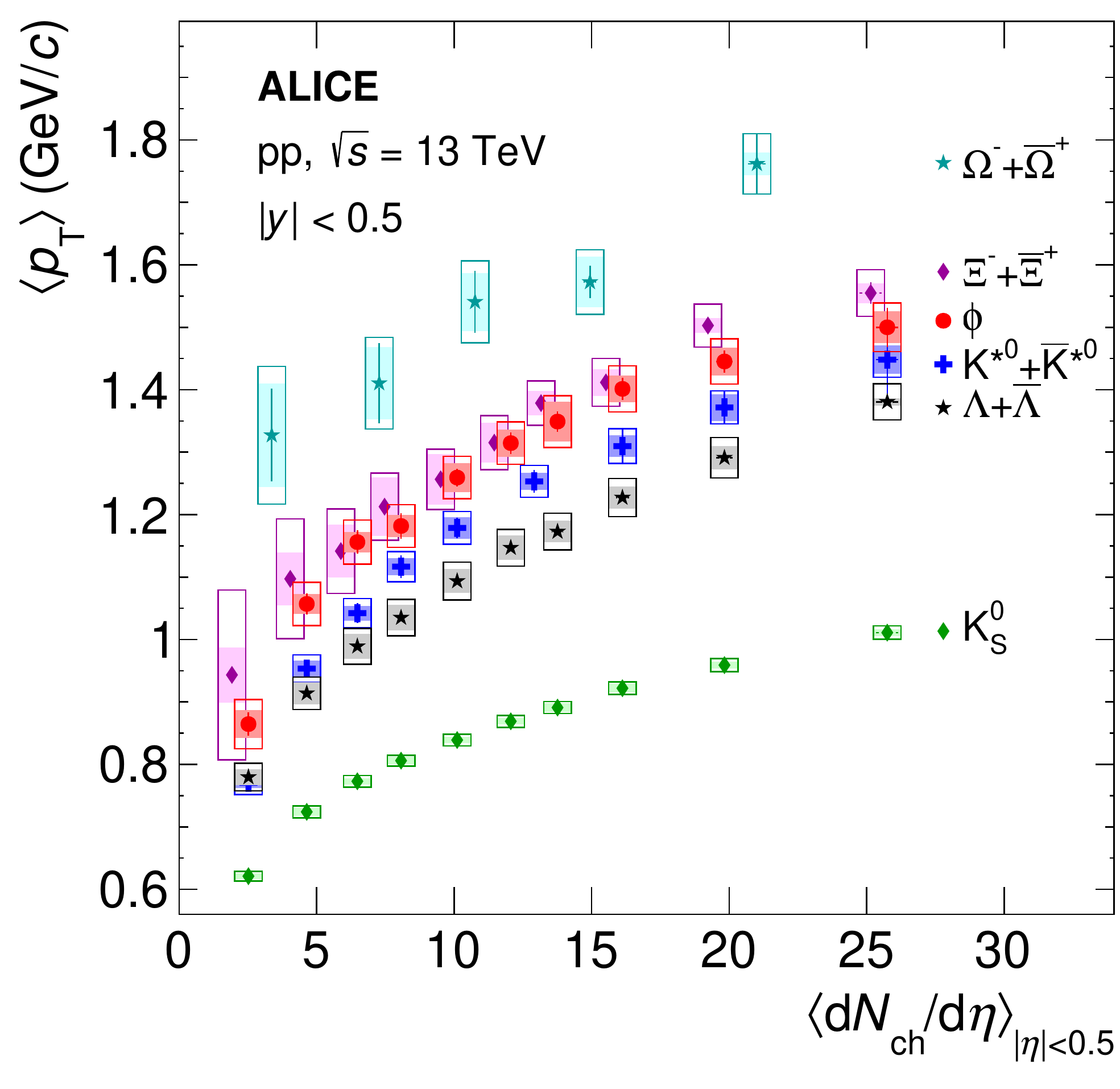}
\hspace{1pc}
\begin{minipage}{18pc}
\vspace{-9cm}
\caption{\label{fig:mpt_others}(Color online) Mean transverse momenta for \ks and \ph are compared with those for \kz, $\Lambda$+$\overline{\Lambda}$, $\Xi^{-}$+$\overline{\Xi}^{+}$, and $\Omega^{-}$+$\overline{\Omega}^{+}$ in pp collisions at \rs as a function of \dnc~\cite{ALICE_strange_pp13}. The values for $\Xi^{-}$+$\overline{\Xi}^{+}$ are shifted horizontally for visibility. Bars represent statistical uncertainties, open boxes represent total systematic uncertainties, and shaded boxes show the systematic uncertainties that are uncorrelated between multiplicity classes.}
\end{minipage}
\end{figure*}

\begin{figure*}
\includegraphics[width=38pc]{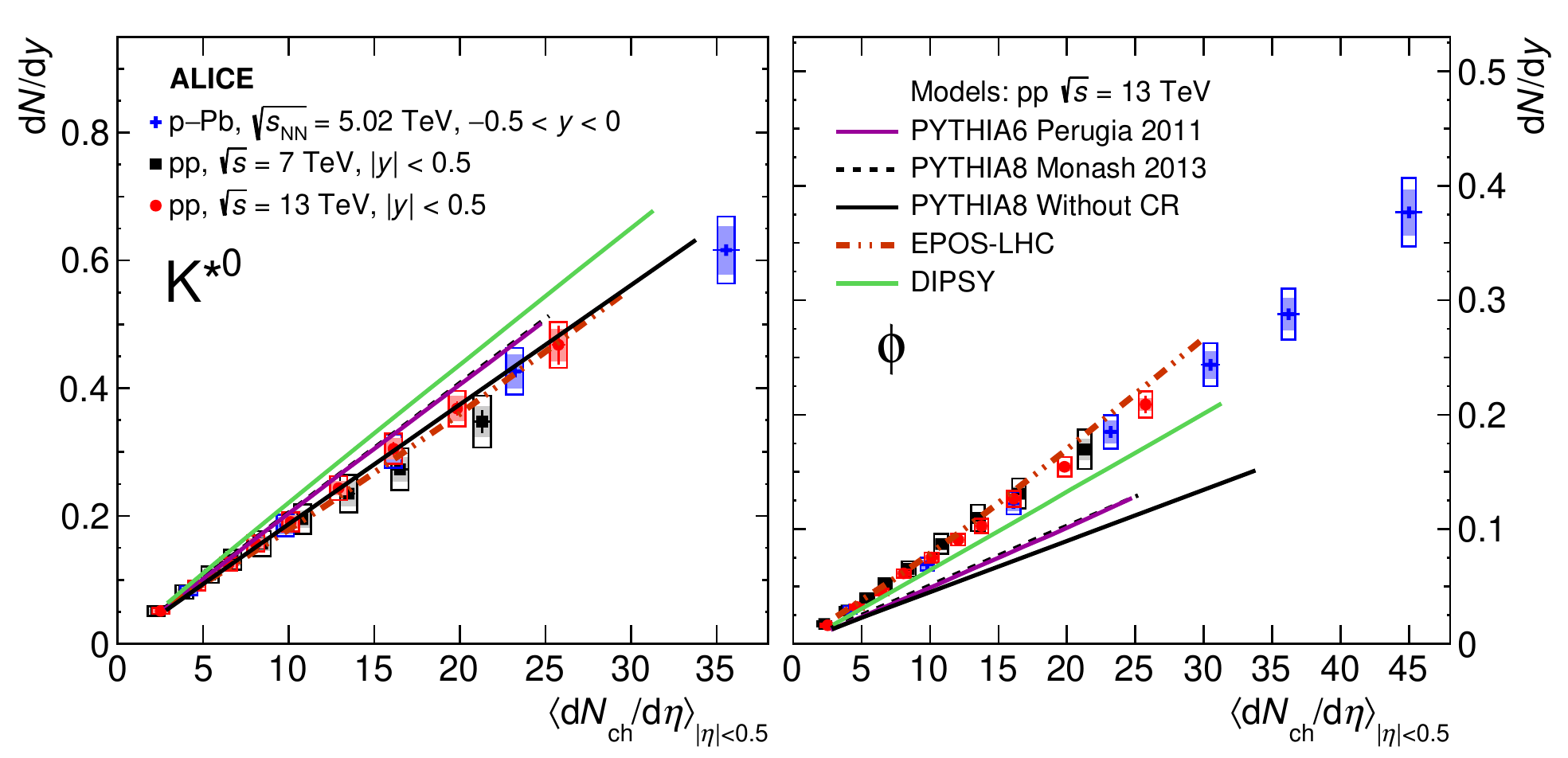}
\caption{\label{fig:yield}(Color online) \ptt-integrated yields \dndy of \ks (average of the particle and antiparticle) and \ph as functions of \dnc. Results are shown for pp collisions at \rs[13] and 7~TeV~\cite{ALICE_LF_pp7}, as well as for \ppb collisions at \rsnn[5.02~TeV]~\cite{ALICE_Kstar_phi_pPb5}. The measurements in pp collisions at \rs are also compared with values from common event generators~\cite{PYTHIA6_Perugia,PYTHIA8_Monash,EPOS_LHC,DIPSY}. Bars represent statistical uncertainties, open boxes represent total systematic uncertainties, and shaded boxes show the systematic uncertainties that are uncorrelated between multiplicity classes.}
\end{figure*}

\begin{figure*}
\includegraphics[width=38pc]{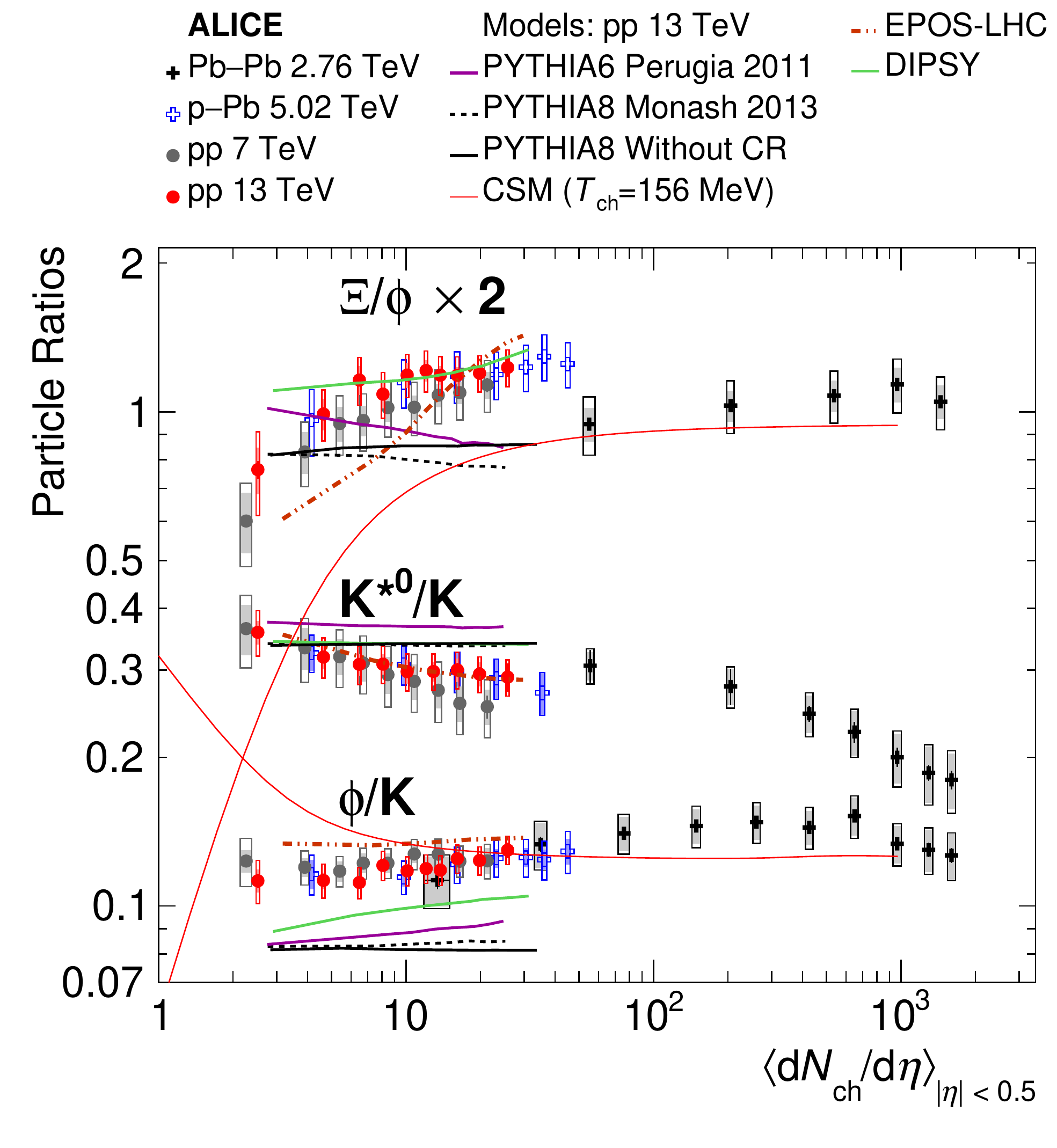}
\caption{\label{fig:ratios}(Color online) Ratios of \ptt-integrated particle yields \ksk, \phik, and \xiphi in pp collisions at \rs as functions of \dnc~\cite{ALICE_strange_pp13}. These measurements are compared with data from \ppb collisions at \rsnn[5.02~TeV]~\cite{ALICE_Xi_Omega_pPb5,ALICE_Kstar_phi_pPb5} and \pb collisions at \rsnn~\cite{ALICE_Kstar_phi_PbPb276,ALICE_Kstar_phi_PbPb276_highpt}, as well as results from common event generators~\cite{PYTHIA6_Perugia,PYTHIA8_Monash,EPOS_LHC,DIPSY} and a Canonical Statistical Model calculation~\cite{ALICE_LF_pp7}.}
\end{figure*}

The \ptt-integrated yields \dndy and mean transverse momenta \mpt are extracted from the \ptt spectra in the different multiplicity classes. For each multiplicity class, the \ph yield is extrapolated to the unmeasured region ($\ptt<0.5$~\gvc) by fitting a L\'{e}vy-Tsallis function~\cite{Tsallis_1988,Wilk_LevyTsallis,STAR_Kstar_AuAu200} to the measured \ptt spectra. For multiplicity class I (X) the extrapolated \ph yield is 12\% (34\%) of the total yield. Uncertainties in \dndy and \mpt are evaluated by varying the fit range and the form of the extrapolation function: Bose-Einstein, Boltzmann, and Boltzmann-Gibbs blast-wave~\cite{BoltzmannGibbsBlastWave} distributions, as well as an exponential in \mT (where $\mT\equiv\sqrt{M^{2}+p_{T}^{2}}$ and $M$ is the mass of the particle). The uncertainty in the total \ph yield due to the extrapolation in class I (X) is 1\% (4.4\%). The \ks is measured down to $\ptt=0$ and no low-\ptt extrapolation is needed. In both cases, the extrapolated yield at high \ptt is negligible.
%For calculations of \dndy and \mpt, the effect of each systematic variation is evaluated simultaneously in the minimum bias event class and the individual multiplicity classes. In this way, we evaluate the degree to which each systematic variation causes a shift that is correlated across multiplicity bins.
The systematic uncertainties on the yield and \mpt are obtained by varying the criteria used in the default analysis. To investigate whether the changes in the yield \dndy and \mpt are correlated across different multiplicity bins, the effect of changing each criterion is simultaneously evaluated for both the minimum bias event class and each individual multiplicity class.
 The multiplicity-correlated and uncorrelated components of the systematic uncertainties are separated, with the latter being plotted as shaded boxes in Figs.~\ref{fig:mpt}-\ref{fig:ratios}.

The mean transverse momenta \mpt for \ks and \ph are shown in Fig.~\ref{fig:mpt} as functions of \dnc and compared with other ALICE measurements and results from model calculations. The \mpt values in pp collisions at \rs[7]~TeV~\cite{ALICE_LF_pp7} and 13~TeV follow approximately the same trend. The \mpt values of \ks and \ph rise slightly faster as a function of \dnc in pp collisions than in \ppb collisions for $\dnc\gtrsim5$; the \mpt values in pp and \ppb collisions both rise faster than those in \pb collisions as discussed in~\cite{ALICE_Kstar_phi_pPb5,ALICE_LF_pp7}. The measured \mpt values are compared with five different model calculations: PYTHIA6 (Perugia 2011 tune)~\cite{PYTHIA6_Perugia}, PYTHIA8 (Monash 2013 tune, both with and without color reconnection)~\cite{PYTHIA8_Monash}, EPOS-LHC~\cite{EPOS_LHC}, and DIPSY~\cite{DIPSY}. PYTHIA8 without color reconnection provides an almost constant \mpt as \dnc increases; this is a very different behavior with respect to the trends measured by ALICE and given by the other model calculations. Turning color reconnection on in PYTHIA8 gives better qualitative agreement with the measurements, although the calculation still somewhat underestimates the \mpt values for hadrons containing strange quarks (\kz, \ks, \ph, $\Lambda$, $\Xi$, and $\Omega$)~\cite{ALICE_strange_pp13}. Color reconnection in PYTHIA8 introduces a flow-like effect, resulting in an increase in \mpt values with increasing multiplicity without assuming the formation of a medium that could flow~\cite{Ortiz_2013}. PYTHIA 6 provides a good description of the \mpt values for \ph, but underestimates \mpt for \ks. The \mpt values predicted by EPOS-LHC are consistent with the measured values for \ph, but slightly below the values for \ks. Among the model results obtained for the present work, EPOS-LHC gives the best agreement with the measured data. DIPSY gives a larger increase in \mpt from low to high \dnc than is actually observed; this discrepancy is greater for the \ph and is also observed for other strange hadrons~\cite{ALICE_strange_pp13}.

The values of \mpt for \ks and \ph are compared with those for \kz and strange baryons in the same collision system in Fig.~\ref{fig:mpt_others}. In central \ada collisions, a mass ordering of the \mpt values is observed; particles with similar masses (e.g., \ks, p, and \ph) have similar \mpt~\cite{ALICE_piKp_PbPb,ALICE_Kstar_phi_pPb5}. This behavior has been interpreted as evidence that radial flow could be a dominant factor in determining the shapes of hadron \ptt spectra in central \ada collisions. However, this mass ordering breaks down for peripheral \pb collisions, as well as \ppb and pp collisions (see Fig. 7 in~\cite{ALICE_Sigmastar_Xistar_pPb5} and measurements reported in~\cite{ALICE_LF_pp7,ALICE_strange_pp13}). In pp collisions at \rs, the \mpt values for \ks and \ph are greater than those for the more massive $\Lambda$ for the same multiplicity classes. The \mpt values for \ph even approach those for $\Xi$, despite the aproximately 30\% larger mass of the $\Xi$. This could be a manifestation of differences between the \ptt spectra of mesons and baryons or different behavior for resonances in comparison to the longer lived particles.
In~\cite{ALICE_LF_pp7}, the Boltzmann-Gibbs blast-wave model was used to predict the \ptt spectra of light-flavor hadrons based on a combined fit of \pix, \kx, and (anti)proton \ptt spectra. This study suggested that strange hadrons (\kz, $\Lambda$, $\Xi$, and $\Omega$) and other light-flavor hadrons might participate in a common radial flow, even in pp collisions, but that \ks and \ph do not follow this common radial expansion (for details of this study, see~\cite{ALICE_LF_pp7}).
%In~\cite{ALICE_LF_pp7}, a combined Boltzmann-Gibbs blast-wave fit was performed of the \ptt spectra of \pix, \kx, and p in high-multiplicity pp collisions at \rs[7~TeV], covering the low-\ptt parts of these spectra ($0.5\leq\ptt(\pix)\leq1$~\gvc, $0.2\leq\ptt(\kx)\leq1.5$~\gvc, and $0.3\leq\ptt(\mathrm{p})\leq3$~\gvc). Using these fit parameters, the blast-wave model was able to describe the \ptt spectra of \kz, $\Lambda$, $\Xi$, and $\Omega$ in pp collisions at \rs[7~TeV] with fair accuracy; \ks and \ph were not well described, however. This suggests that \pix, K, p, and the strange baryons might participate in common collective motion, from which \ks and \ph are excluded.
The same behavior could result in the violation of mass ordering for \mpt seen at \rs. A deviation of the \mpt values of short-lived resonances above the trend for other hadrons could in principle be explained by re-scattering of the resonance-decay daughters during the hadronic phase of the collision, which is expected to be most important at low \ptt~\cite{Bleicher_Stoecker}. However, the strongest re-scattering phenomena occur in central \ada collisions, where no deviation from mass ordering is observed. In addition, such effects would be stronger for the shorter lived \ks than for the \ph, which decays predominantly outside the hadronic phase (even in central \ada collisions) and should be minimally affected by re-scattering. On the other hand, the observed violation of mass ordering could be due to differences between baryon and meson \ptt spectra. Baryon-to-meson ratios such as $\mathrm{p}/\pion$ and $\Lambda/\kz$ are observed~\cite{ALICE_LF_pp7,ALICE_piKpLambda_pPb5} to be enhanced at intermediate \ptt ($\sim3$~\gvc), even in pp and \ppb collisions, while similar enhancement is not observed in meson-to-meson ratios like $\mathrm{K}/\pion$. Differences between baryons and mesons have also been observed in the \mT spectra of hadrons measured at RHIC energies~\cite{STAR_strange_pp200,PHENIX_neutral_mesons_pp200}. For $\mT\gtrsim1$~\gvc, meson \mT spectra follow one common trend, while baryons follow a different, more steeply falling trend as a function of \mT. Such differences between the shapes of baryon and meson spectra may result in mesons having larger \mpt values than baryons with comparable masses. The breakdown of mass ordering, with $\langle\ptt(\mathrm{p})\rangle<\langle\ptt(\ks)\rangle\approx\langle\ptt(\Lambda)\rangle<\langle\ptt(\ph)\rangle\approx\langle\ptt(\Xi)\rangle$, is a common feature of the models shown in Fig.~\ref{fig:mpt}.
%The breakdown of mass ordering is a common feature of the models shown in Fig.~\ref{fig:mpt}, with $\langle\ptt(\mathrm{p})\rangle<\langle\ptt(\ks)\rangle\approx\langle\ptt(\Lambda)\rangle$ and $\langle\ptt(\Lambda)\rangle<\langle\ptt(\ph)\rangle\approx\langle\ptt(\Xi)\rangle$ for all five calculations.
This behavior may be a consequence of hadron production via fragmentation at high \ptt or \mT; meson formation requires only the production of a quark-antiquark pair, while baryon formation requires a diquark-antidiquark pair~\cite{STAR_strange_pp200}.

The \ptt-integrated yields of \ks and \ph are shown in Fig.~\ref{fig:yield} as functions of \dnc. For both particles, \dndy exhibits an approximately linear increase with increasing \dnc. Results for pp collisions at \rs[7] and 13~TeV and for \ppb collisions at \rsnn[5.02~TeV] follow approximately the same trends. This indicates that, for a given multiplicity, \ks and \ph production does not depend on the collision system or energy. Similar results are seen for strange hadrons~\cite{ALICE_strange_pp13}. The \dndy values are also compared with those obtained from the same models studied for the discussion of \mpt. For the \ks, EPOS-LHC and PYTHIA8 without color reconnection give the best descriptions, the other PYTHIA calculations exhibit fair agreement with the measured data, and DIPSY tends to overestimate the \ks yields. The \ph yields tend to be slightly overestimated by EPOS-LHC and slightly underestimated by DIPSY, while the PYTHIA calculations underestimate the \ph yields by about 40\%. The selected PYTHIA tunes also underestimate the yields of $\Lambda$, $\Xi$, and $\Omega$ by similar factors~\cite{ALICE_strange_pp13}. For these baryons, the EPOS-LHC description becomes less accurate with increasing strangeness content; DIPSY describes the $\Lambda$ and $\Xi$ yields well, but underestimates the yields of $\Omega$~\cite{ALICE_strange_pp13}.

\begin{sloppypar} The ratios of the \ptt-integrated particle yields \ksk and \phik are shown in Fig.~\ref{fig:ratios} as functions of \dnc. Within their uncertainties the ratios in pp collisions at \rs[7] and 13~TeV and in \ppb collisions at \rsnn[5.02~TeV] are consistent for similar values of \dnc. There is a hint of a decrease in \ksk with increasing \dnc in all three collision systems; for pp collisions at \rs the \ksk ratio in the highest multiplicity class is below the low-multiplicity value at the 2.3~$\sigma$ level. The decrease in \ksk in central \pb collisions~\cite{ALICE_Kstar_phi_pPb5,ALICE_Kstar_phi_PbPb276,ALICE_Kstar_phi_PbPb276_highpt} has been attributed to re-scattering of the \ks decay products in the hadronic phase of the collision~\cite{EPOS_resonances_PbPb}. It remains an open question whether a decrease in pp collisions could be caused by the same mechanism. EPOS-LHC provides the best description of the \ksk ratio in pp collisions at \rs. PYTHIA and DIPSY tend to overestimate the ratio for large multiplicities and do not reproduce the apparent decrease with increasing \dnc. The \phik ratio also follows a similar trend in the three collision systems. It is fairly constant as a function of \dnc, although there is an apparent small increase with \dnc from the lowest multiplicities up to $\dnc\approx400$. EPOS-LHC somewhat overestimates the \phik ratio, but is closer to the measured values than PYTHIA, which significantly underestimates \phik. While PYTHIA6 and DIPSY underestimate the \phik ratio, both results exhibit small increases with increasing multiplicity, which is qualitatively similar to the measured trend. In addition, Fig.~\ref{fig:ratios} also includes the results of a canonical statistical model (CSM) calculation~\cite{ALICE_LF_pp7} with a chemical freeze-out temperature of 156~MeV; this calculation does not describe the behavior of the measured \phik ratio for the \dnc range spanned by the ALICE pp measurements.\end{sloppypar}
%, the only model shown that includes re-scattering effects,

In addition to comparing the yields of \ph and kaons, it may be instructive to compare $\Xi$ and \ph. These two particles contain the same number of strange valence (anti)quarks: \ph is a s$\bar{\mathrm{s}}$ bound state and $\Xi$ contains two strange valence quarks. However, $\Xi$ would be subject to canonical suppression, unlike the strangeness-neutral \ph. Figure~\ref{fig:ratios} also shows the \xiphi ratio in pp, \ppb, and \pb collisions. The ratio increases with increasing \dnc for low-multiplicity collisions and is then fairly constant for a wide range of multiplicities: from high-multiplicity pp and \ppb collisions to central \pb collisions. The decrease in \xiphi with decreasing \dnc for low multiplicities could be interpreted as evidence of canonical suppression in small systems; the canonical statistical model predicts a decrease in the \xiphi ratio with decreasing \dnc that is qualitatively similar to the measured data. However, canonical suppression would also result in an increase in the \phik ratio with decreasing \dnc, which is not observed. Given that $\Xi$ and K have different numbers of strange valence (anti)quarks, it is expected that $\Xi$ would be more affected by canonical suppression~\cite{ALICE_LF_pp7}. It will be interesting to extend the study of the \phik ratio to lower multiplicities to test if there is any increase in this ratio due to canonical suppression of kaon yields. Even in the absence of canonical suppression, the multiplicity evolution of the \xiphi and \phik ratios suggests that the \ph meson behaves as if it had between 1 and 2 units of strangeness: i.e., $\Xi$ is enhanced more than \ph, which is (possibly) enhanced more than K. In addition, there are indications of increases in the p/$\pion$ and $\Lambda/\kz$ ratios with increasing \dnc~\cite{ALICE_LF_pp7,ALICE_strange_pp13} which are qualitatively similar to the increase in \xiphi, but smaller in magnitude. This suggests that baryon-meson differences (e.g., baryon suppression or meson enhancement) might be a contributing factor, but not the only reason, for the low-multiplicity behavior of the \xiphi ratio. EPOS-LHC, which includes core-corona effects, gives an increasing trend in \xiphi with increasing \dnc, although the values of the ratio and its flattening at high multiplicity are not particularly well described. In contrast, PYTHIA gives a constant or decreasing value of \xiphi with increasing \dnc, which is inconsistent with the observed trend. DIPSY, which includes rope hadronization, describes the \xiphi ratio over a wide \dnc range, only failing to describe the decrease in the ratio with decreasing multiplicity for the lowest \dnc values.

\begin{figure*}
\includegraphics[width=38pc]{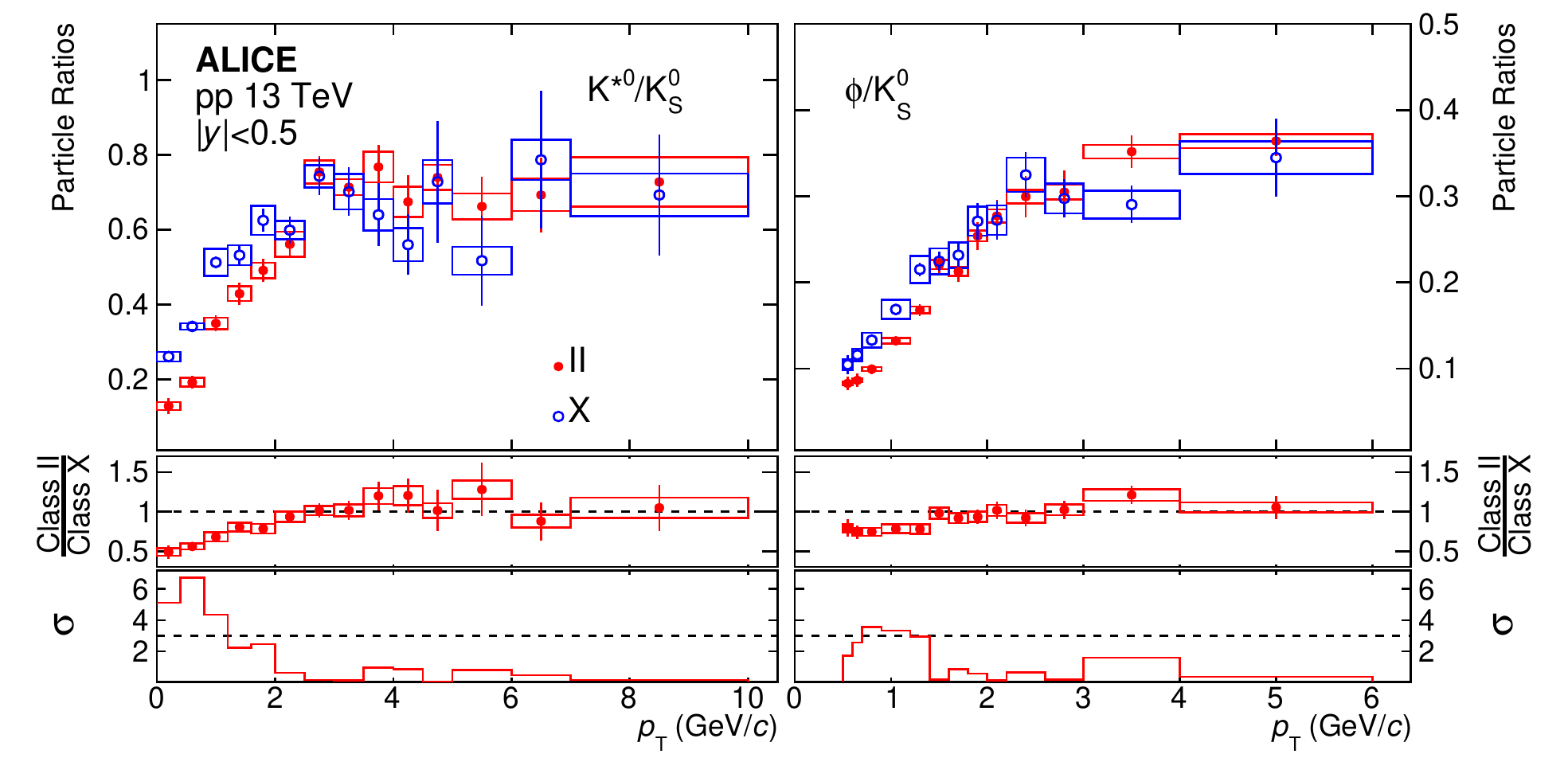}
\caption{\label{fig:ratios_pt}(Color online) Ratios of particle yields \kskz and \phikz as functions of \ptt~\cite{ALICE_strange_pp13} for low (X) and high (II) multiplicity classes. The middle panels show the double ratios: the measurements in class II divided by those in class X. The significance of the deviations of the double ratios from unity is plotted in the lower panels, with dashed lines indicating a deviation at the 3 $\sigma$ level. Bars represent the statistical uncertainties, while boxes represent the part of the systematic uncertainty that is uncorrelated between multiplicity classes II and X.}
\end{figure*}

The \ptt dependence of the particle ratios \kskz and \phikz is shown in Fig.~\ref{fig:ratios_pt} for low and high multiplicity classes (X and II, respectively). Both ratios increase at low \ptt and saturate for $\ptt\gtrsim2.5$~\gvc; however, for $\ptt\lesssim2.5$~\gvc the \kskz and \phikz ratios in the high multiplicity class (II) are less than in the lowest multiplicity class (X). This behavior is qualitatively consistent with observations in \pb collisions at \rsnn[2.76~TeV]~\cite{ALICE_Kstar_phi_PbPb276_highpt}, where the \ksk and \phik ratios at low \ptt in central collisions are lower compared to pp collisions. The decrease in the low-\ptt \ksk ratio in central \pb collisions with respect to pp collisions is larger than the decrease in the \phik ratio, which could be expected due to the presence of re-scattering effects. To quantify the decrease in these particle ratios in pp collisions at \rs, the middle panels of Fig.~\ref{fig:ratios_pt} show the double ratios: the high-multiplicity values divided by the low multiplicity values. The double ratios are consistent with unity for $\ptt\gtrsim2.5$~\gvc, which suggests a common evolution of the \ptt spectra for these three mesons. However for $\ptt\lesssim2.5$~\gvc, the suppression of the \kskz ratio from low to high-multiplicity collisions is greater than the suppression of the \phikz ratio. This is quantified in the lower panels of Fig.~\ref{fig:ratios_pt}, where the significance of the deviations of the double ratios from unity is shown. For $\ptt<1.2$~\gvc, the \kskz double ratio deviates from unity by 4--6.6 times its standard deviation, while the \phikz double ratio deviates from unity at about the 3~$\sigma$ level for $0.6<\ptt<1.4$~\gvc.
%For $\ptt<1.2$~\gvc, the average deviation from unity is 5.4$\sigma$ for \kskz and 2.7$\sigma$ for \phikz.
This difference may be a hint of re-scattering in small collision systems.

\section{Conclusions\label{sec:conclusions}}

The ALICE Collaboration has reported measurements of the \ks and \ph mesons at midrapidity in pp collisions at \rs in multiplicity classes. The results have many qualitative similarities to those reported for longer lived strange hadrons in~\cite{ALICE_strange_pp13} and may be evidence for collective behavior in small collision systems. The slopes of the \ptt spectra of \ks and \ph are observed to increase with increasing multiplicity for $\ptt\lesssim4$~\gvc, which is qualitatively similar to the collective radial expansion observed in \pb collisions, but can also be explained through color reconnection. In contrast, the shapes of the \ptt spectra are the same for all multiplicity classes at high \ptt. Both the \ptt-integrated yields and the mean transverse momenta increase with increasing charged-particle multiplicity at midrapidity, with approximately linear increases for the yields. It appears that, for a given multiplicity value, the yields of these particles are independent of collision system and energy, while the \mpt values follow different trends for different collision systems. The mass ordering of the \mpt values observed in central \pb collisions is violated in pp collisions, with the \ks and \ph mesons having greater \mpt than baryons with similar masses. The EPOS-LHC model describes the multiplicity dependence of the yields and \mpt fairly well for pp collisions at \rs. There are hints that the yields of \ks may be reduced, particularly at low \ptt and high multiplicity, by re-scattering of its decay daughters in a short-lived hadron-gas phase in pp collisions; similar behavior is observed in \pb collisions.
The yields of the \ph meson evolve similarly to particles with 1 and 2 units of open strangeness. The \phik and \xiphi ratios are both fairly constant, exhibiting only slow increases over wide multiplicity ranges, although the \xiphi ratio decreases with decreasing \dnc for the lowest multiplicity pp and \ppb collisions. In high-multiplicity pp and \ppb collisions, these ratios reach values observed in central \pb collisions. This multiplicity evolution is not consistent with simple descriptions of canonical suppression, but is qualitatively described by the DIPSY model, which includes rope hadronization effects.
%The \xiphi ratio is approximately constant over a wide multiplicity range in pp, \ppb, and \pb collisions, but exhibits a decrease with decreasing multiplicity for the lowest multiplicity pp and \ppb collisions. The multiplicity evolution of the \xiphi ratio is fairly well described by the DIPSY model, which includes rope hadronization effects.
These new measurements of the \ph provide further constraints for theoretical models of strangeness production in small collision systems.

%%%%% acknowledgements
\newenvironment{acknowledgement}{\relax}{\relax}
\begin{acknowledgement}
\section*{Acknowledgements}
% Version: 2019-10-08

The ALICE Collaboration would like to thank all its engineers and technicians for their invaluable contributions to the construction of the experiment and the CERN accelerator teams for the outstanding performance of the LHC complex.
The ALICE Collaboration gratefully acknowledges the resources and support provided by all Grid centres and the Worldwide LHC Computing Grid (WLCG) collaboration.
The ALICE Collaboration acknowledges the following funding agencies for their support in building and running the ALICE detector:
A. I. Alikhanyan National Science Laboratory (Yerevan Physics Institute) Foundation (ANSL), State Committee of Science and World Federation of Scientists (WFS), Armenia;
Austrian Academy of Sciences, Austrian Science Fund (FWF): [M 2467-N36] and Nationalstiftung f\"{u}r Forschung, Technologie und Entwicklung, Austria;
Ministry of Communications and High Technologies, National Nuclear Research Center, Azerbaijan;
Conselho Nacional de Desenvolvimento Cient\'{\i}fico e Tecnol\'{o}gico (CNPq), Financiadora de Estudos e Projetos (Finep), Funda\c{c}\~{a}o de Amparo \`{a} Pesquisa do Estado de S\~{a}o Paulo (FAPESP) and Universidade Federal do Rio Grande do Sul (UFRGS), Brazil;
Ministry of Education of China (MOEC) , Ministry of Science \& Technology of China (MSTC) and National Natural Science Foundation of China (NSFC), China;
Ministry of Science and Education and Croatian Science Foundation, Croatia;
Centro de Aplicaciones Tecnol\'{o}gicas y Desarrollo Nuclear (CEADEN), Cubaenerg\'{\i}a, Cuba;
Ministry of Education, Youth and Sports of the Czech Republic, Czech Republic;
The Danish Council for Independent Research | Natural Sciences, the VILLUM FONDEN and Danish National Research Foundation (DNRF), Denmark;
Helsinki Institute of Physics (HIP), Finland;
Commissariat \`{a} l'Energie Atomique (CEA), Institut National de Physique Nucl\'{e}aire et de Physique des Particules (IN2P3) and Centre National de la Recherche Scientifique (CNRS) and R\'{e}gion des  Pays de la Loire, France;
Bundesministerium f\"{u}r Bildung und Forschung (BMBF) and GSI Helmholtzzentrum f\"{u}r Schwerionenforschung GmbH, Germany;
General Secretariat for Research and Technology, Ministry of Education, Research and Religions, Greece;
National Research, Development and Innovation Office, Hungary;
Department of Atomic Energy Government of India (DAE), Department of Science and Technology, Government of India (DST), University Grants Commission, Government of India (UGC) and Council of Scientific and Industrial Research (CSIR), India;
Indonesian Institute of Science, Indonesia;
Centro Fermi - Museo Storico della Fisica e Centro Studi e Ricerche Enrico Fermi and Istituto Nazionale di Fisica Nucleare (INFN), Italy;
Institute for Innovative Science and Technology , Nagasaki Institute of Applied Science (IIST), Japanese Ministry of Education, Culture, Sports, Science and Technology (MEXT) and Japan Society for the Promotion of Science (JSPS) KAKENHI, Japan;
Consejo Nacional de Ciencia (CONACYT) y Tecnolog\'{i}a, through Fondo de Cooperaci\'{o}n Internacional en Ciencia y Tecnolog\'{i}a (FONCICYT) and Direcci\'{o}n General de Asuntos del Personal Academico (DGAPA), Mexico;
Nederlandse Organisatie voor Wetenschappelijk Onderzoek (NWO), Netherlands;
The Research Council of Norway, Norway;
Commission on Science and Technology for Sustainable Development in the South (COMSATS), Pakistan;
Pontificia Universidad Cat\'{o}lica del Per\'{u}, Peru;
Ministry of Science and Higher Education and National Science Centre, Poland;
Korea Institute of Science and Technology Information and National Research Foundation of Korea (NRF), Republic of Korea;
Ministry of Education and Scientific Research, Institute of Atomic Physics and Ministry of Research and Innovation and Institute of Atomic Physics, Romania;
Joint Institute for Nuclear Research (JINR), Ministry of Education and Science of the Russian Federation, National Research Centre Kurchatov Institute, Russian Science Foundation and Russian Foundation for Basic Research, Russia;
Ministry of Education, Science, Research and Sport of the Slovak Republic, Slovakia;
National Research Foundation of South Africa, South Africa;
Swedish Research Council (VR) and Knut \& Alice Wallenberg Foundation (KAW), Sweden;
European Organization for Nuclear Research, Switzerland;
Suranaree University of Technology (SUT), National Science and Technology Development Agency (NSDTA) and Office of the Higher Education Commission under NRU project of Thailand, Thailand;
Turkish Atomic Energy Agency (TAEK), Turkey;
National Academy of  Sciences of Ukraine, Ukraine;
Science and Technology Facilities Council (STFC), United Kingdom;
National Science Foundation of the United States of America (NSF) and United States Department of Energy, Office of Nuclear Physics (DOE NP), United States of America.    %%%%%%% done by webmaster team
\end{acknowledgement}

%%%%%%%% Bibliography (In case of using bibtex generate the bbl requested by arXiv)
\bibliography{refs}{}
\bibliographystyle{utphys}   % Remember we use title in the biblio

%%%%%%%%% appendix with author list
\newpage
\appendix
\section{The ALICE Collaboration}
\label{app:collab}
% Collaboration: CERN-LHC-ALICE
% Generation Date is 2019-10-08

% How to use:
%%%%%%%%% appendix with author list
%\appendix
%\section{The ALICE Collaboration}
%\label{app:collab}
%\input{Alice_Authorslist_XXXX-Axx-XX.tex}
\begingroup
\small
\begin{flushleft}
S.~Acharya\Irefn{org141}\And 
D.~Adamov\'{a}\Irefn{org94}\And 
A.~Adler\Irefn{org74}\And 
J.~Adolfsson\Irefn{org80}\And 
M.M.~Aggarwal\Irefn{org99}\And 
G.~Aglieri Rinella\Irefn{org33}\And 
M.~Agnello\Irefn{org30}\And 
N.~Agrawal\Irefn{org10}\textsuperscript{,}\Irefn{org53}\And 
Z.~Ahammed\Irefn{org141}\And 
S.~Ahmad\Irefn{org16}
% ...
\And
\renewcommand\labelenumi{\textsuperscript{\theenumi}~}

\section*{Affiliation notes}
\renewcommand\theenumi{\roman{enumi}}
\begin{Authlist}
\item \Adef{org*}Deceased
\item \Adef{orgI}Dipartimento DET del Politecnico di Torino, Turin, Italy
\item \Adef{orgII}M.V. Lomonosov Moscow State University, D.V. Skobeltsyn Institute of Nuclear, Physics, Moscow, Russia
\item \Adef{orgIII}Department of Applied Physics, Aligarh Muslim University, Aligarh, India
\item \Adef{orgIV}Institute of Theoretical Physics, University of Wroclaw, Poland
\end{Authlist}

\section*{Collaboration Institutes}
\renewcommand\theenumi{\arabic{enumi}~}
\begin{Authlist}
\item \Idef{org1}A.I. Alikhanyan National Science Laboratory (Yerevan Physics Institute) Foundation, Yerevan, Armenia
\item \Idef{org2}Bogolyubov Institute for Theoretical Physics, National Academy of Sciences of Ukraine, Kiev, Ukraine
\item \Idef{org3}Bose Institute, Department of Physics  and Centre for Astroparticle Physics and Space Science (CAPSS), Kolkata, India
\item \Idef{org4}Budker Institute for Nuclear Physics, Novosibirsk, Russia
\item \Idef{org5}California Polytechnic State University, San Luis Obispo, California, United States
\item \Idef{org6}Central China Normal University, Wuhan, China
\item \Idef{org7}Centre de Calcul de l'IN2P3, Villeurbanne, Lyon, France
\item \Idef{org8}Centro de Aplicaciones Tecnol\'{o}gicas y Desarrollo Nuclear (CEADEN), Havana, Cuba
\item \Idef{org9}Centro de Investigaci\'{o}n y de Estudios Avanzados (CINVESTAV), Mexico City and M\'{e}rida, Mexico
\item \Idef{org10}Centro Fermi - Museo Storico della Fisica e Centro Studi e Ricerche ``Enrico Fermi', Rome, Italy
\item \Idef{org11}Chicago State University, Chicago, Illinois, United States
\item \Idef{org12}China Institute of Atomic Energy, Beijing, China
\item \Idef{org13}Comenius University Bratislava, Faculty of Mathematics, Physics and Informatics, Bratislava, Slovakia
\item \Idef{org14}COMSATS University Islamabad, Islamabad, Pakistan
\item \Idef{org15}Creighton University, Omaha, Nebraska, United States
\item \Idef{org16}Department of Physics, Aligarh Muslim University, Aligarh, India
\item \Idef{org17}Department of Physics, Pusan National University, Pusan, Republic of Korea
\item \Idef{org18}Department of Physics, Sejong University, Seoul, Republic of Korea
\item \Idef{org19}Department of Physics, University of California, Berkeley, California, United States
\item \Idef{org20}Department of Physics, University of Oslo, Oslo, Norway
\item \Idef{org21}Department of Physics and Technology, University of Bergen, Bergen, Norway
\item \Idef{org22}Dipartimento di Fisica dell'Universit\`{a} 'La Sapienza' and Sezione INFN, Rome, Italy
\item \Idef{org23}Dipartimento di Fisica dell'Universit\`{a} and Sezione INFN, Cagliari, Italy
\item \Idef{org24}Dipartimento di Fisica dell'Universit\`{a} and Sezione INFN, Trieste, Italy
\item \Idef{org25}Dipartimento di Fisica dell'Universit\`{a} and Sezione INFN, Turin, Italy
\item \Idef{org26}Dipartimento di Fisica e Astronomia dell'Universit\`{a} and Sezione INFN, Bologna, Italy
\item \Idef{org27}Dipartimento di Fisica e Astronomia dell'Universit\`{a} and Sezione INFN, Catania, Italy
\item \Idef{org28}Dipartimento di Fisica e Astronomia dell'Universit\`{a} and Sezione INFN, Padova, Italy
\item \Idef{org29}Dipartimento di Fisica `E.R.~Caianiello' dell'Universit\`{a} and Gruppo Collegato INFN, Salerno, Italy
\item \Idef{org30}Dipartimento DISAT del Politecnico and Sezione INFN, Turin, Italy
\item \Idef{org31}Dipartimento di Scienze e Innovazione Tecnologica dell'Universit\`{a} del Piemonte Orientale and INFN Sezione di Torino, Alessandria, Italy
\item \Idef{org32}Dipartimento Interateneo di Fisica `M.~Merlin' and Sezione INFN, Bari, Italy
\item \Idef{org33}European Organization for Nuclear Research (CERN), Geneva, Switzerland
\item \Idef{org34}Faculty of Electrical Engineering, Mechanical Engineering and Naval Architecture, University of Split, Split, Croatia
\item \Idef{org35}Faculty of Engineering and Science, Western Norway University of Applied Sciences, Bergen, Norway
\item \Idef{org36}Faculty of Nuclear Sciences and Physical Engineering, Czech Technical University in Prague, Prague, Czech Republic
\item \Idef{org37}Faculty of Science, P.J.~\v{S}af\'{a}rik University, Ko\v{s}ice, Slovakia
\item \Idef{org38}Frankfurt Institute for Advanced Studies, Johann Wolfgang Goethe-Universit\"{a}t Frankfurt, Frankfurt, Germany
\item \Idef{org39}Fudan University, Shanghai, China
\item \Idef{org40}Gangneung-Wonju National University, Gangneung, Republic of Korea
\item \Idef{org41}Gauhati University, Department of Physics, Guwahati, India
\item \Idef{org42}Helmholtz-Institut f\"{u}r Strahlen- und Kernphysik, Rheinische Friedrich-Wilhelms-Universit\"{a}t Bonn, Bonn, Germany
\item \Idef{org43}Helsinki Institute of Physics (HIP), Helsinki, Finland
\item \Idef{org44}High Energy Physics Group,  Universidad Aut\'{o}noma de Puebla, Puebla, Mexico
\item \Idef{org45}Hiroshima University, Hiroshima, Japan
\item \Idef{org46}Hochschule Worms, Zentrum  f\"{u}r Technologietransfer und Telekommunikation (ZTT), Worms, Germany
\item \Idef{org47}Horia Hulubei National Institute of Physics and Nuclear Engineering, Bucharest, Romania
\item \Idef{org48}Indian Institute of Technology Bombay (IIT), Mumbai, India
\item \Idef{org49}Indian Institute of Technology Indore, Indore, India
\item \Idef{org50}Indonesian Institute of Sciences, Jakarta, Indonesia
\item \Idef{org51}INFN, Laboratori Nazionali di Frascati, Frascati, Italy
\item \Idef{org52}INFN, Sezione di Bari, Bari, Italy
\item \Idef{org53}INFN, Sezione di Bologna, Bologna, Italy
\item \Idef{org54}INFN, Sezione di Cagliari, Cagliari, Italy
\item \Idef{org55}INFN, Sezione di Catania, Catania, Italy
\item \Idef{org56}INFN, Sezione di Padova, Padova, Italy
\item \Idef{org57}INFN, Sezione di Roma, Rome, Italy
\item \Idef{org58}INFN, Sezione di Torino, Turin, Italy
\item \Idef{org59}INFN, Sezione di Trieste, Trieste, Italy
\item \Idef{org60}Inha University, Incheon, Republic of Korea
\item \Idef{org61}Institut de Physique Nucl\'{e}aire d'Orsay (IPNO), Institut National de Physique Nucl\'{e}aire et de Physique des Particules (IN2P3/CNRS), Universit\'{e} de Paris-Sud, Universit\'{e} Paris-Saclay, Orsay, France
\item \Idef{org62}Institute for Nuclear Research, Academy of Sciences, Moscow, Russia
\item \Idef{org63}Institute for Subatomic Physics, Utrecht University/Nikhef, Utrecht, Netherlands
\item \Idef{org64}Institute of Experimental Physics, Slovak Academy of Sciences, Ko\v{s}ice, Slovakia
\item \Idef{org65}Institute of Physics, Homi Bhabha National Institute, Bhubaneswar, India
\item \Idef{org66}Institute of Physics of the Czech Academy of Sciences, Prague, Czech Republic
\item \Idef{org67}Institute of Space Science (ISS), Bucharest, Romania
\item \Idef{org68}Institut f\"{u}r Kernphysik, Johann Wolfgang Goethe-Universit\"{a}t Frankfurt, Frankfurt, Germany
\item \Idef{org69}Instituto de Ciencias Nucleares, Universidad Nacional Aut\'{o}noma de M\'{e}xico, Mexico City, Mexico
\item \Idef{org70}Instituto de F\'{i}sica, Universidade Federal do Rio Grande do Sul (UFRGS), Porto Alegre, Brazil
\item \Idef{org71}Instituto de F\'{\i}sica, Universidad Nacional Aut\'{o}noma de M\'{e}xico, Mexico City, Mexico
\item \Idef{org72}iThemba LABS, National Research Foundation, Somerset West, South Africa
\item \Idef{org73}Jeonbuk National University, Jeonju, Republic of Korea
\item \Idef{org74}Johann-Wolfgang-Goethe Universit\"{a}t Frankfurt Institut f\"{u}r Informatik, Fachbereich Informatik und Mathematik, Frankfurt, Germany
\item \Idef{org75}Joint Institute for Nuclear Research (JINR), Dubna, Russia
\item \Idef{org76}Korea Institute of Science and Technology Information, Daejeon, Republic of Korea
\item \Idef{org77}KTO Karatay University, Konya, Turkey
\item \Idef{org78}Laboratoire de Physique Subatomique et de Cosmologie, Universit\'{e} Grenoble-Alpes, CNRS-IN2P3, Grenoble, France
\item \Idef{org79}Lawrence Berkeley National Laboratory, Berkeley, California, United States
\item \Idef{org80}Lund University Department of Physics, Division of Particle Physics, Lund, Sweden
\item \Idef{org81}Nagasaki Institute of Applied Science, Nagasaki, Japan
\item \Idef{org82}Nara Women{'}s University (NWU), Nara, Japan
\item \Idef{org83}National and Kapodistrian University of Athens, School of Science, Department of Physics , Athens, Greece
\item \Idef{org84}National Centre for Nuclear Research, Warsaw, Poland
\item \Idef{org85}National Institute of Science Education and Research, Homi Bhabha National Institute, Jatni, India
\item \Idef{org86}National Nuclear Research Center, Baku, Azerbaijan
\item \Idef{org87}National Research Centre Kurchatov Institute, Moscow, Russia
\item \Idef{org88}Niels Bohr Institute, University of Copenhagen, Copenhagen, Denmark
\item \Idef{org89}Nikhef, National institute for subatomic physics, Amsterdam, Netherlands
\item \Idef{org90}NRC Kurchatov Institute IHEP, Protvino, Russia
\item \Idef{org91}NRC «Kurchatov Institute»  - ITEP, Moscow, Russia
\item \Idef{org92}NRNU Moscow Engineering Physics Institute, Moscow, Russia
\item \Idef{org93}Nuclear Physics Group, STFC Daresbury Laboratory, Daresbury, United Kingdom
\item \Idef{org94}Nuclear Physics Institute of the Czech Academy of Sciences, \v{R}e\v{z} u Prahy, Czech Republic
\item \Idef{org95}Oak Ridge National Laboratory, Oak Ridge, Tennessee, United States
\item \Idef{org96}Ohio State University, Columbus, Ohio, United States
\item \Idef{org97}Petersburg Nuclear Physics Institute, Gatchina, Russia
\item \Idef{org98}Physics department, Faculty of science, University of Zagreb, Zagreb, Croatia
\item \Idef{org99}Physics Department, Panjab University, Chandigarh, India
\item \Idef{org100}Physics Department, University of Jammu, Jammu, India
\item \Idef{org101}Physics Department, University of Rajasthan, Jaipur, India
\item \Idef{org102}Physikalisches Institut, Eberhard-Karls-Universit\"{a}t T\"{u}bingen, T\"{u}bingen, Germany
\item \Idef{org103}Physikalisches Institut, Ruprecht-Karls-Universit\"{a}t Heidelberg, Heidelberg, Germany
\item \Idef{org104}Physik Department, Technische Universit\"{a}t M\"{u}nchen, Munich, Germany
\item \Idef{org105}Politecnico di Bari, Bari, Italy
\item \Idef{org106}Research Division and ExtreMe Matter Institute EMMI, GSI Helmholtzzentrum f\"ur Schwerionenforschung GmbH, Darmstadt, Germany
\item \Idef{org107}Rudjer Bo\v{s}kovi\'{c} Institute, Zagreb, Croatia
\item \Idef{org108}Russian Federal Nuclear Center (VNIIEF), Sarov, Russia
\item \Idef{org109}Saha Institute of Nuclear Physics, Homi Bhabha National Institute, Kolkata, India
\item \Idef{org110}School of Physics and Astronomy, University of Birmingham, Birmingham, United Kingdom
\item \Idef{org111}Secci\'{o}n F\'{\i}sica, Departamento de Ciencias, Pontificia Universidad Cat\'{o}lica del Per\'{u}, Lima, Peru
\item \Idef{org112}St. Petersburg State University, St. Petersburg, Russia
\item \Idef{org113}Stefan Meyer Institut f\"{u}r Subatomare Physik (SMI), Vienna, Austria
\item \Idef{org114}SUBATECH, IMT Atlantique, Universit\'{e} de Nantes, CNRS-IN2P3, Nantes, France
\item \Idef{org115}Suranaree University of Technology, Nakhon Ratchasima, Thailand
\item \Idef{org116}Technical University of Ko\v{s}ice, Ko\v{s}ice, Slovakia
\item \Idef{org117}Technische Universit\"{a}t M\"{u}nchen, Excellence Cluster 'Universe', Munich, Germany
\item \Idef{org118}The Henryk Niewodniczanski Institute of Nuclear Physics, Polish Academy of Sciences, Cracow, Poland
\item \Idef{org119}The University of Texas at Austin, Austin, Texas, United States
\item \Idef{org120}Universidad Aut\'{o}noma de Sinaloa, Culiac\'{a}n, Mexico
\item \Idef{org121}Universidade de S\~{a}o Paulo (USP), S\~{a}o Paulo, Brazil
\item \Idef{org122}Universidade Estadual de Campinas (UNICAMP), Campinas, Brazil
\item \Idef{org123}Universidade Federal do ABC, Santo Andre, Brazil
\item \Idef{org124}University of Cape Town, Cape Town, South Africa
\item \Idef{org125}University of Houston, Houston, Texas, United States
\item \Idef{org126}University of Jyv\"{a}skyl\"{a}, Jyv\"{a}skyl\"{a}, Finland
\item \Idef{org127}University of Liverpool, Liverpool, United Kingdom
\item \Idef{org128}University of Science and Techonology of China, Hefei, China
\item \Idef{org129}University of South-Eastern Norway, Tonsberg, Norway
\item \Idef{org130}University of Tennessee, Knoxville, Tennessee, United States
\item \Idef{org131}University of the Witwatersrand, Johannesburg, South Africa
\item \Idef{org132}University of Tokyo, Tokyo, Japan
\item \Idef{org133}University of Tsukuba, Tsukuba, Japan
\item \Idef{org134}Universit\'{e} Clermont Auvergne, CNRS/IN2P3, LPC, Clermont-Ferrand, France
\item \Idef{org135}Universit\'{e} de Lyon, Universit\'{e} Lyon 1, CNRS/IN2P3, IPN-Lyon, Villeurbanne, Lyon, France
\item \Idef{org136}Universit\'{e} de Strasbourg, CNRS, IPHC UMR 7178, F-67000 Strasbourg, France, Strasbourg, France
\item \Idef{org137}Universit\'{e} Paris-Saclay Centre d'Etudes de Saclay (CEA), IRFU, D\'{e}partment de Physique Nucl\'{e}aire (DPhN), Saclay, France
\item \Idef{org138}Universit\`{a} degli Studi di Foggia, Foggia, Italy
\item \Idef{org139}Universit\`{a} degli Studi di Pavia, Pavia, Italy
\item \Idef{org140}Universit\`{a} di Brescia, Brescia, Italy
\item \Idef{org141}Variable Energy Cyclotron Centre, Homi Bhabha National Institute, Kolkata, India
\item \Idef{org142}Warsaw University of Technology, Warsaw, Poland
\item \Idef{org143}Wayne State University, Detroit, Michigan, United States
\item \Idef{org144}Westf\"{a}lische Wilhelms-Universit\"{a}t M\"{u}nster, Institut f\"{u}r Kernphysik, M\"{u}nster, Germany
\item \Idef{org145}Wigner Research Centre for Physics, Budapest, Hungary
\item \Idef{org146}Yale University, New Haven, Connecticut, United States
\item \Idef{org147}Yonsei University, Seoul, Republic of Korea
\end{Authlist}
\endgroup
  %%%%%%% done by webmaster team
\end{document}